\documentclass{article}
\pdfoutput=1
\usepackage[activate={true,nocompatibility},final,tracking=true,kerning=true,spacing=true,factor=1100,stretch=10,shrink=10]{microtype}

\usepackage{arxiv}
\usepackage{nameref}
\usepackage[font=small,labelfont={bf}, figurename=Figure, tablename=Table]{caption}

\usepackage[hidelinks]{hyperref}
\hypersetup{
  colorlinks   = true, 
  urlcolor     = esalightblue, 
  linkcolor    = esablue, 
  citecolor   = esablue 
}
\usepackage{amsmath}
\usepackage{amssymb}
\usepackage{bbold}
\usepackage[utf8]{inputenc} 
\usepackage[T1]{fontenc}    
\usepackage{hyperref}       
\usepackage{url}            
\usepackage{booktabs}       
\usepackage{amsfonts}       
\usepackage{nicefrac}       
\usepackage{lipsum}
\usepackage{graphicx}
\graphicspath{ {./images/} }
\usepackage[nameinlink, capitalise]{cleveref}
\usepackage{xcolor}
\definecolor{esablue}{RGB}{0,57,158}
\definecolor{esalightblue}{RGB}{0,155,219}
\definecolor{esared}{RGB}{150,1,54}
\newcommand{\review}[1]{#1}

\newcommand{\rvec}[2]{\pmb r^{#2}_{#1}}

\newcommand{\rvecxn}[1]{r^{#1}_{x}}
\newcommand{\rvecyn}[1]{r^{#1}_{y}}

\newcommand{\nvec}[1]{\pmb n_{#1}}
\newcommand{\nvecp}[1]{\pmb n'_{#1}}

\newcommand{\mvec}[1]{\pmb m_{#1}}
\newcommand{\mvecp}[1]{\pmb m'_{#1}}

\newcommand{\origin}[1]{\pmb o_{#1}}

\newcommand{\beam}{l_\text{b}}
\newcommand{\lever}{\frac{\beam}{2}}

\newcommand{\polar}[1]{\pmb e^{#1}}
\newcommand{\polarx}[1]{e^{#1}_x}
\newcommand{\polary}[1]{e^{#1}_y}

\newcommand{\springp}[1]{\pmb s^+_{#1}}

\newcommand{\springpyn}{s^+_{y}}
\newcommand{\springm}[1]{\pmb s^-_{#1}}

\newcommand{\springmxn}{s^-_{x}}
\newcommand{\springmyn}{s^-_{y}}

\newcommand{\weirdvec}[1]{\pmb k_{#1}}

\newcommand{\pA}{\mathrm{A}}
\newcommand{\pP}{\mathrm{P}}
\newcommand{\pB}{\mathrm{B}}
\newcommand{\pC}{\mathrm{C}}
\newcommand{\pAp}{\mathrm{A'}}
\newcommand{\pPp}{\mathrm{P'}}
\newcommand{\pBp}{\mathrm{B'}}

\newcommand{\beamAngle}[1]{\varphi_{#1}}
\newcommand{\beamAnglep}[1]{\varphi'_{#1}}

\newcommand{\beamAngley}{\varphi_y}
\newcommand{\beamAngleyp}{\varphi'_y}
\newcommand{\beamAnglez}{\varphi_z}
\newcommand{\beamAnglezp}{\varphi'_z}

\newcommand{\beamAngleyt}[1]{\varphi_{y;#1}}
\newcommand{\beamAngleypt}[1]{\varphi'_{y;#1}}
\newcommand{\beamAnglezt}[1]{\varphi_{z;#1}}
\newcommand{\beamAnglezpt}[1]{\varphi'_{z;#1}}

\newcommand{\leverAngle}[1]{\theta_{#1}}
\newcommand{\leverAnglex}{\theta_x}
\newcommand{\leverAnglez}{\theta_z}
\newcommand{\leverAnglexp}{\theta'_x}
\newcommand{\leverAnglezp}{\theta'_z}
\newcommand{\leverAnglep}[1]{\theta'_{#1}}
\newcommand{\invBeamAnglep}[1]{\gamma'_{#1}}

\newcommand{\stopleft}{\alpha'_\mathrm{L}}
\newcommand{\stopright}{\alpha'_\mathrm{R}}
\newcommand{\sbotleft}{\alpha_\mathrm{L}}
\newcommand{\sbotright}{\alpha_\mathrm{R}}

\newcommand{\rot}[2]{R_{#1}(#2)}

\newcommand{\spherical}[2]{\pmb \xi^{#1,#2}}
\newcommand{\sphericaly}[2]{\pmb \epsilon^{#1,#2}}

\newcommand{\invPolar}[1]{E\left( #1 \right)}
\newcommand{\invSph}[1]{\Xi\left( #1 \right)}
\newcommand{\invSphy}[1]{\mathcal{E}\left( #1 \right)}

\newcommand{\xd}[1]{u^{x}_{#1}}
\newcommand{\yd}[1]{u^{y}_{#1}}
\newcommand{\phid}[1]{u^{\varphi}_{#1}}
\newcommand{\Fx}[1]{F^{x}_{#1}}
\newcommand{\Fy}[1]{F^{y}_{#1}}
\newcommand{\Mphi}[1]{M^{\varphi}_{#1}}

\newcommand{\dof}{df}
\newcommand{\totcoord}{\mathcal{P}}
\newcommand{\lattcoord}{\mathcal{R}}
\newcommand{\totfunc}{f_\mathrm{T}}
\newcommand{\femfunc}{f_\mathrm{DS}}
\newcommand{\focal}{\pmb f^\mathrm{tgt}}
\newcommand{\rayfunc}{f_\mathrm{O}}

\usepackage[giveninits=true,sorting=none,citestyle=numeric-comp]{biblatex}
\begin{document}

\title{\Large{Continuous Design and Reprogramming of Totimorphic Structures for Space Applications}}

\author{
  Dominik Dold$^{1,2,\dagger}$, Amy Thomas$^{1,3}$, Nicole Rosi$^{1}$, Jai Grover$^{1}$, Dario Izzo$^{1}$\\[2pt]
  \textit{$^1$ \small{Advanced Concepts Team,  European Space Agency, European Space Research and Technology Centre, The Netherlands}}\\
  \textit{$^2$ \small{Faculty of Mathematics,  University of Vienna, Austria}}\\
  \textit{$^3$ \small{Delft University of Technology, The Netherlands}}\\
  \textit{$^\dagger$\small{dominik.dold@univie.ac.at}}\\
\vspace{-7mm}
}

\maketitle
\begin{abstract}
Recently, a class of mechanical lattices with reconfigurable, zero-stiffness structures has been proposed, called Totimorphic lattices.
In this work, we introduce a computational framework that enables continuous reprogramming of a Totimorphic lattice's effective properties, such as mechanical and optical behaviour, through geometric changes alone, demonstrated using computer simulations.
Our approach is differentiable and guarantees valid Totimorphic configurations throughout the optimisation process, providing not only target states with desired properties but also continuous trajectories in configuration space that connect them.
This enables reprogrammable structures in which actuators are controlled via automatic differentiation on an objective-dependent cost function, continuously adapting the lattice to achieve a given goal.
We focus on deep space applications, where harsh and resource-constrained environments demand solutions that combine flexibility, efficiency, and autonomy. As proof of concept, we present two scenarios: a reprogrammable disordered lattice material and a space telescope mirror with adjustable focal length. 
The introduced framework is adaptable to a wide range of Totimorphic designs and objectives, providing a lightweight model for endowing physical systems with autonomous self-configuration and self-repair capabilities.
\end{abstract}


\section{Introduction}

\begin{figure*}[h!]
    \centering
    \includegraphics[width=\textwidth]{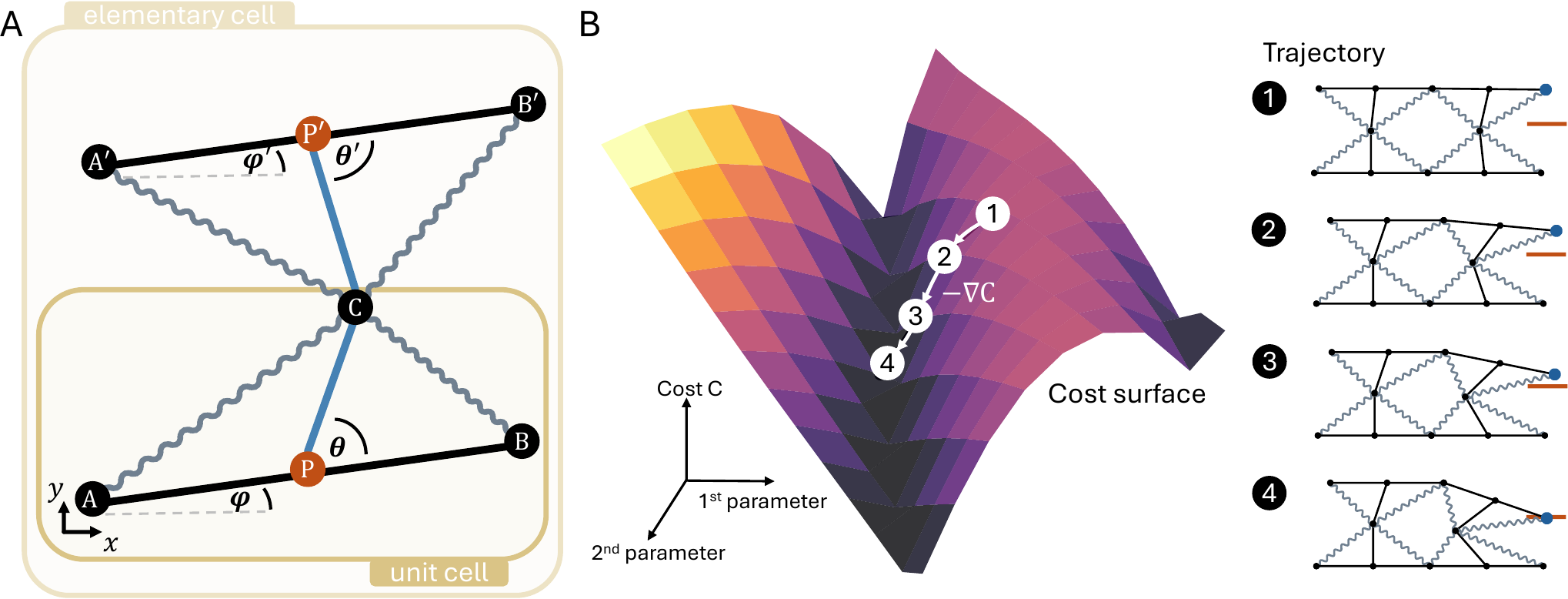}
	\caption{Illustration of Totimorphic lattices reconfigured using automatic differentiation. \textbf{(A)} Totimorphic unit cell. For tiling, we use the shown elementary cell as the basic building block in this work. \textbf{(B)} Actuation of generalized coordinates (here: beam and lever angles) is defined as gradient descent on an objective-dependent cost function $C$. This way, both a lattice configuration with a desired property is found as well as the trajectory connecting it to the initial configuration. The surface plot represents the cost for different configurations. An example trajectory following the gradient $-\nabla C$ is shown in white (from 1 to 4), and the corresponding lattice configurations to the right. For illustration purposes, the task here is to move the upper right corner of the lattice (blue dot) to the height of the red line. In realistic scenarios, the landscape might contain holes, i.e., configurations that break the structure.}
	\label{fig:Intro}
\end{figure*}

Self-assembling and reconfigurable space infrastructure has the potential to reduce deployment costs and risks while increasing mission flexibility and long-term reusability.
A variety of concepts for in-orbit assembly have lately been proposed, including formation flight \cite{mettler2005large,ayre2005self}, CubeSat swarms that dock into space mirrors \cite{pirat2022toward}, autonomously assembling tiles for habitat construction \cite{ayre2005self,ekblaw2021self}, origami-based deployment of spacecraft components \cite{trease2013accommodating,zirbel2014deployment}, and robotic systems for ground-based assembly on other celestial bodies \cite{gregg2024ultralight}.
However, most of these approaches focus solely on deployment and offer little or no capability for post-deployment reconfiguration, and are therefore unable to adapt to new tasks, repair damage, or respond to shifting environmental conditions autonomously.
Recent advances such as Mechanical Neural Networks \cite{lee2022mechanical} and Trimorph origami patterns \cite{liu2022triclinic} begin to explore programmable materials with reprogrammable properties, but remain limited to discrete configurations or narrow ranges of tunable properties.

Nature offers powerful examples of how disordered geometry alone can endow structures with exceptional mechanical properties \cite{wegst2015bioinspired}.
Prominent examples are bones, plant stems, dragonfly wings, coral, and radiolarians \cite{gibson2010cellular}, and the concept is further found in human-engineered metamaterials \cite{surjadi2019mechanical,meza2014strong,xiao2020active,xia2022responsive}.
Inspired by recent work on inversely designing irregular mechanical lattices \cite{aranguren2018designing,torres2019bone,ross2021using,maurizi2022inverse,dold2023differentiable,shumilin2023method,zheng2023unifying,jadhav2024generative}, we propose a new concept for reconfigurable space infrastructure with a high degree of programmability: continuously inverse-designed Totimorphic lattices. 

Totimorphic lattices are a recently introduced type of reconfigureable mechanical lattice \cite{chaudhary2021totimorphic}. 
Their triangular unit cell consists of a beam (\cref{fig:Intro}A, points $\pA$ to $\pB$), a lever attached at the beam’s midpoint (points $\pP$ to $\pC$), and two zero-length springs (springs where the spring force is proportional to the total length of the spring) that connect the lever tip to the beam ends (wiggly lines). \review{The beam-lever connection and connections between unit cells use ball (pin) joints}. Because of this arrangement, the spring forces always balance without exerting torque on the lever. As a result, each unit cell — and any lattice assembled from them — is neutrally stable: it can reconfigure freely under external actuation, yet remains stationary when unloaded.

In previous work \cite{chaudhary2021totimorphic}, target lattice shapes were obtained through constrained optimisation, with beam and lever lengths enforced as constraints. This approach has two main drawbacks: constraints are only satisfied approximately, and just the final configuration is produced, not the trajectory through configuration space to get there.
\review{In \cite{thomas37totimorphic}, we demonstrated shape-morphing, e.g., from a flat surface sheet into a half-cylinder, using a marching cell method, which however struggled with maintaining lattice integrity and could become stuck while morphing between shapes}.
Here we address both issues by introducing generalised coordinates that absorb the geometric constraints into the parameterisation itself. The resulting mapping from generalised to physical lattice coordinates is differentiable, allowing us to use automatic differentiation on task-specific cost functions. This way, Totimorphic lattices are reconfigured continuously along valid trajectories until the cost function is minimised (\cref{fig:Intro}B).

\begin{figure*}[h!]
    \centering
    \includegraphics[width=0.65\textwidth]{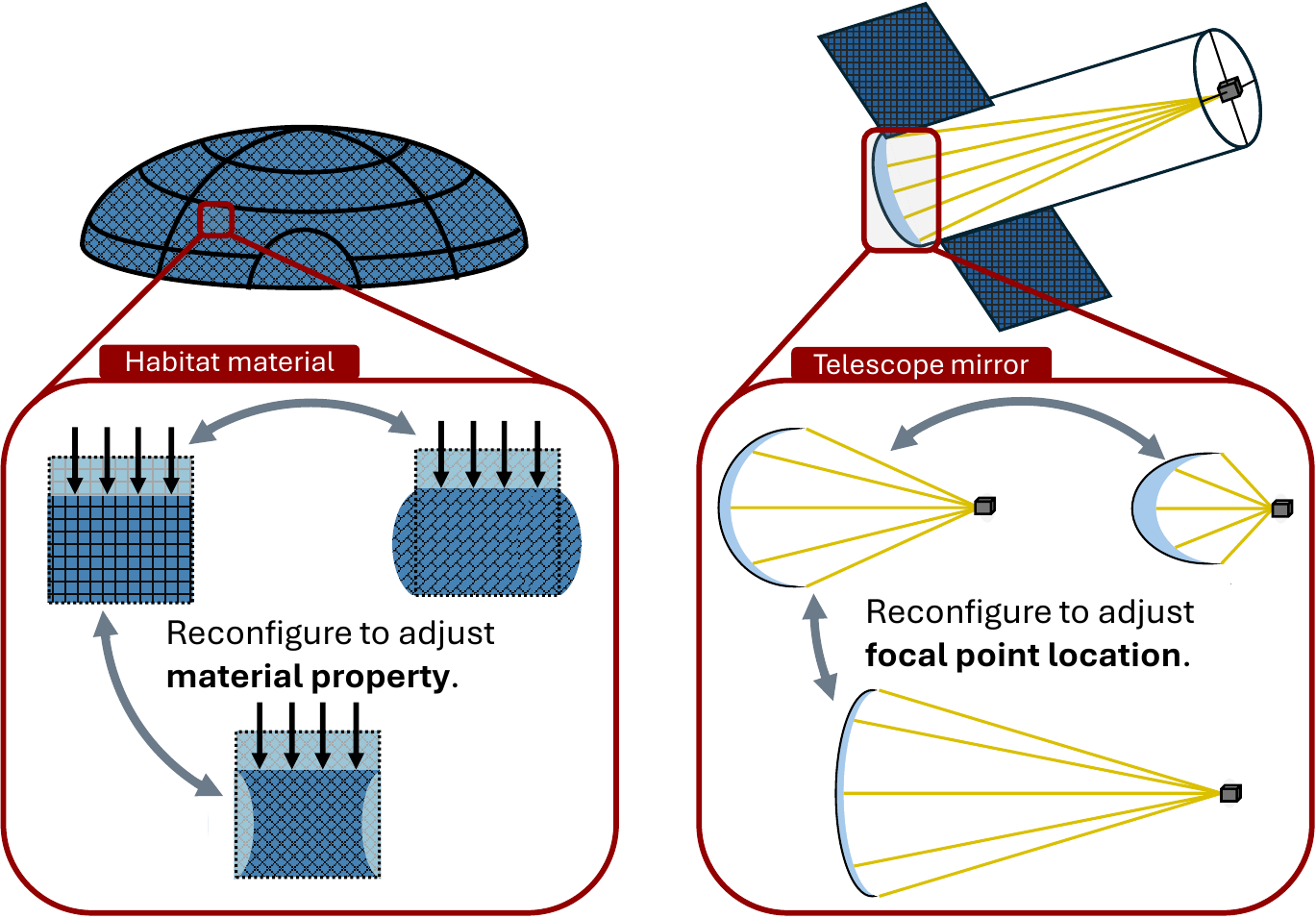}
	\caption{Totimorphic lattices as building blocks of, e.g., space habitats, spacecraft components, and tools
with reprogrammable properties. In this work, we specifically look into lattice structures that can alter their mechanical properties, i.e., how they behave under stress. In addition, we explore the concept of using Totimorphic lattices as the backbone of large-scale space infrastructures such as space telescopes -- not only to ease their deployment, but to equip them with reprogrammable properties as well. For example, we demonstrate a mirror element that can adjust its surface shape to alter the location of its focal point.}
	\label{fig:Concept}
\end{figure*}

We validate our framework in the space domain using simulations, adding to the growing number of concepts of self-constructing and reconfigurable designs. 
As a first demonstration, we show that Totimorphic lattices can continuously tune their effective mechanical properties, making them feasible as building blocks for adaptive structures with tunable stiffness or Poisson’s ratio (Fig.~\ref{fig:Concept}, left). 
As a second, we explore their use in large-scale infrastructure by simulating a reconfigurable space mirror capable of adapting its focal length and repairing structural damage (Fig.~\ref{fig:Concept}, right).

In the following, we first introduce the generalised coordinates and differentiable reconfiguration framework, followed by the two proof of concept space applications.
Mathematical details on all models, proofs, and simulation details are found in \cref{sec:methods}.
Simulation code is available online on GitHub~\cite{github}.

\section{Results}


\subsection{Degrees of freedom in Totimorphic lattices}\label{methods:2d}

To model Totimorphic lattices, we absorb the beam and lever length / angle constraints into the parameters used to describe the lattice's configuration. The minimum number of parameters needed to uniquely capture the lattice is given by its degrees of freedom $\dof$. In general, the degrees of freedom of an elementary cell in a Totimorphic lattice are determined by three factors: the freely moving points characterising the cell itself, the beam/lever constraints, and points that overlap with other cells in the lattice -- with the latter two factors reducing the number of degrees of freedom.
In the following, we construct a lattice cell by cell to identify how a cell's degrees of freedom depend on where and when it is added to the lattice.
For simplicity, we showcase this for lattices in two dimensions:
\begin{enumerate}
	\item The first elementary cell that we add is not constrained by any neighbouring cells (Fig.~\ref{fig:Model}A1).
	It is characterised by five points: $\pA_0$, $\pB_0$, $\pC_0$, $\pA'_0$, and $\pB'_0$.
	In two dimensions, this yields ten parameters (two for each point).
	However, due to the beam/lever constraints, i.e. the constant lengths of the four beams, there are only six degrees of freedom (each length constraint reduces the degrees of freedom by one), and so we only require six parameters to fully describe the elementary cell.
	\item Adding another elementary cell above the first one, the points $\pA_1$ and $\pB_1$ are already determined, $\pA_1 = \pA'_0$ and $\pB_1 = \pB'_0$ (Fig.~\ref{fig:Model}A2).
	Thus, we have six parameters (point $\pC_1$, $\pA'_1$ and $\pB'_1$), which leaves us with three degrees of freedom for this elementary cell when accounting for the three remaining length constraints (two levers and one beam).
	The same is true when adding an elementary cell below the first one.
	\item Adding another elementary cell to the right of the first one, both points $\pA_2$ and $\pA'_2$ are already determined, $\pA_2 = \pB_0$ and $\pA'_2 = \pB'_0$ (Fig.~\ref{fig:Model}A3).
	Since all four length constraints (two levers and two beams) still have to be enforced, we are left with only two degrees of freedom in this case.
	The same is true when adding an elementary cell to the left of the first one.
	\item Adding an elementary cell above and to the right of two other cells, the points $\pA_3$, $\pB_3$ and $\pA'_3$ are already determined, $\pA_3 = \pB'_0$, $\pB_3 = \pB'_2$, $\pA'_3 = \pB'_1$ (Fig.~\ref{fig:Model}A4).
	Since three length constraints still have to be satisfied, we are left with only one degree of freedom (four from the remaining points minus the three length constraints).
    This is in general true when adding an elementary cell such that two of its adjacent sides are connected to other elementary cells.
\end{enumerate}
Following the above logic, we find that for a lattice in two dimensions, the first placed elementary cell has six degrees of freedom, elementary cells to the left and right of the first one have two each, cells above and below have three each, and all other elementary cells have only one.
For a lattice with $C$ columns and $R$ rows, this results in $df_\mathrm{2D} = 2 + C + 2R + C\cdot R$ degrees of freedom in total.
Applying the same logic to a lattice in three dimensions yields eleven, five, six, and three degrees of freedom, respectively, or $df_\mathrm{3D} = 3 + 2C + 3R + 3 C\cdot R$ in total.

\subsection{Angle-based generalised coordinates}\label{methods:3d}

\begin{figure*}[t!]
    \centering
    \includegraphics[width=0.8\textwidth]{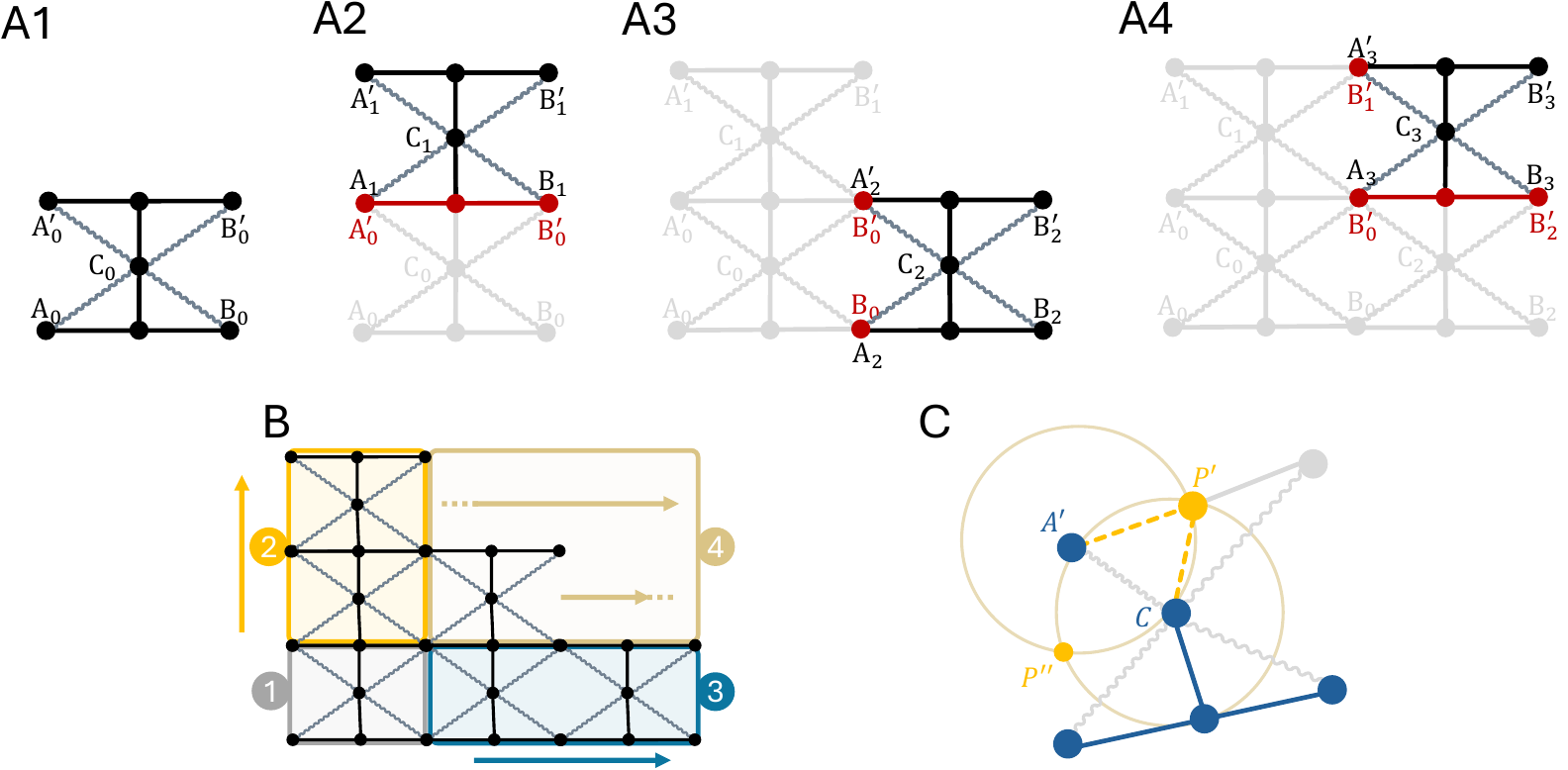}
	\caption{Parametrizing Totimorphic lattices. \textbf{(A)} When connecting elementary cells to an existing lattice, overlapping elements are shared with already existing cells (shown in red). Thus, the number of degrees of freedom of the newly added cell is reduced. \textbf{(B)} Iterative construction of a lattice made of Totimorphic cells. \textbf{(C)} If all points except one in an elementary cell are predetermined (shown in blue), the remaining point is obtained from the beam and lever length constraints. This is illustrated here for the case where $\pAp$ and $\pC$ are given. Since $\pPp$ is connected via a beam and lever to these points, it has to be a distance of $\frac{1}{2}\beam$ away from both. The solution to this problem is given by the intersection points of two circles, or equivalently, by a quadratic equation. Two points $\pPp$ and $\mathrm{P}''$ satisfy the constraints of a Totimorphic lattice, however, the latter would reverse the order of $\pAp$ and $\pBp$ and is therefore not a valid physical solution.}
	\label{fig:Model}
\end{figure*}

In this work, we parameterise all elementary cells using beam and lever angles and a vector indicating the location of the whole lattice in space. 
In addition, we construct a lattice by placing the bottom left elementary cell first and adding all other elementary cells row by row.
We briefly illustrate this process in two dimensions:
\begin{enumerate}
	\item As the first step, we place the bottom left corner elementary cell of the lattice, which has six degrees of freedom (Fig.~\ref{fig:Model}B, box 1).
	The six parameters we choose to describe this elementary cell are the location of the lattice and the orientation of the bottom and top beams and levers.
    \item Then we add the first column, i.e. we stack up elementary cells on top of the first one (Fig.~\ref{fig:Model}B, box 2). 
    Each elementary cell in this column has three degrees of freedom.
    Here, we choose the orientation of bottom and top lever as well as the orientation of the upper beam as parameters.
    \item We then add the first row, i.e. we string together elementary cells to the right of the first one (Fig.~\ref{fig:Model}B, box 3).
    These elementary cells only have two degrees of freedom, which we choose to be the orientation of the bottom beam and lever. 
    In this case, the midpoint of the upper beam ($\pPp$) is given by the (analytic) solution of a quadratic equation representing the beam/lever constraints (Fig.~\ref{fig:Model}C).
	\item Then we fill in the remaining elementary cells row-wise from left to right (Fig.~\ref{fig:Model}B, box 4). All these elementary cells have only one degree of freedom, which we choose to be represented by the orientation of the lower lever.
    The midpoint of the upper beam is again given by the solution of a quadratic equation.
\end{enumerate}
For a detailed mathematical description, see \cref{sec:methodslattice2D}. 
Although we parametrise the lattice using beam and lever angles here, many other parametrisations are possible.
From the quadratic equations, we further derive upper and lower bounds for the lever angles, allowing levers to be pushed or dragged by other lattice elements.
For lattices in two dimensions, additional conditions are derived to avoid configurations with overlapping lattice elements.
Models of lattices in three dimensions are constructed analogously.
For details, see \cref{sec:methodslattice3D}.

\subsection{Differentiable model of Totimorphic lattices}

With this approach, a Totimorphic lattice is represented by a set of $\dof$ parameters $\totcoord$, $|\ \totcoord\ | = \dof$, which in our case is the location of the lattice's bottom left corner in space as well as the angles of levers and beams in the lattice ($\varphi$, $\theta$, and primed variants in \cref{fig:Intro}A) -- although some cells are constrained by neighbouring cells and therefore represented by fewer angles, as described in the previous section.
From this set of parameters, there exists a differentiable function $\totfunc$ which maps to the points of each elementary cell in the lattice, $\lattcoord = \totfunc(\totcoord)$, where $\lattcoord$ is a set of vectors containing the coordinates of each point in the lattice (beam and lever ends, i.e., points $\pA$, $\pB$, $\pP$, $\pC$, and primed variants). We call $\totfunc$ the Totimorphic model in the following.
Invalid lattice configurations are given by values of the generalised coordinates $\totcoord$ for which no solution of $\totfunc$ exists, which can be checked using simple criteria.
Although this approach can be easily applied to any Totimorphic lattice -- in fact, the calculations required to map from generalised coordinates to physical coordinates are always of the same type -- we illustrate it for a specific tiling used in \cite{chaudhary2021totimorphic} where the lattice is formed from hour-glass shaped cells (made from two Totimorphic unit cells), see Fig.~\ref{fig:Intro}A. We call these cells \textit{elementary cells} in the remainder to distinguish them from unit cells.
We derive generalised coordinates for this tiling both in two and three-dimensional space (see \cref{sec:methods}).

To change the configuration of the lattice to one with some desired property (e.g. a desired stiffness), we minimise an objective-dependent cost function that measures how well the current configuration satisfies the desired property.
In most cases, we express the cost function using the physical coordinates obtained from the Totimorphic model.
What is actually being optimised, however, are the parameters $\totcoord$ used to characterise the lattice.
By tracking intermediate configurations, we obtain the trajectory through parameter space connecting the initial and the final lattice configuration.
Each of these configurations is a valid Totimorphic lattice, meaning that the reconfiguration is achieved only through changes in the beam and lever angles (realised via joints), without ever deforming or damaging beams and levers.

In this work, we exclusively use automatic differentiation (i.e. gradient descent) as our optimisation algorithm of choice to guarantee scalability and continuously connected trajectories.
In other terms: for real-world prototypes, the control of the actuators (rotation in the joints) is given by the negative gradient of the cost function.

\subsection{First proof of concept: continuous inverse design of lattice structures}

As a first proof of concept, we explore reconfigurable lattice structures (on the length scale of cm) with adjustable mechanical properties.
We calculate the mechanical properties of a Totimorphic lattice by simulating a compression experiment using the direct stiffness method (see \cref{methods:DS} for details).
In a compression experiment, the lattice is glued between surfaces and slowly compressed.
During compression, the response of the lattice to this load is measured, for example, the displacement of nodes in the lattice -- from which mechanical properties such as effective stiffness and Poisson's ratio are derived.
Here, we restrict the work to the effective Poisson's ratio, which measures how a material expands on average horizontally when compressed vertically.
A positive Poisson's ratio (like honeycomb tiles) means the material widens, while a negative one means it narrows (also known as auxetic).
A Poisson's ratio of zero consequently means that the material does not change its width during compression (see \cref{methods:Poisson} for details).

\subsubsection{Direct stiffness method}
 
\begin{figure*}[b!]
    \centering
    \includegraphics[width=\textwidth]{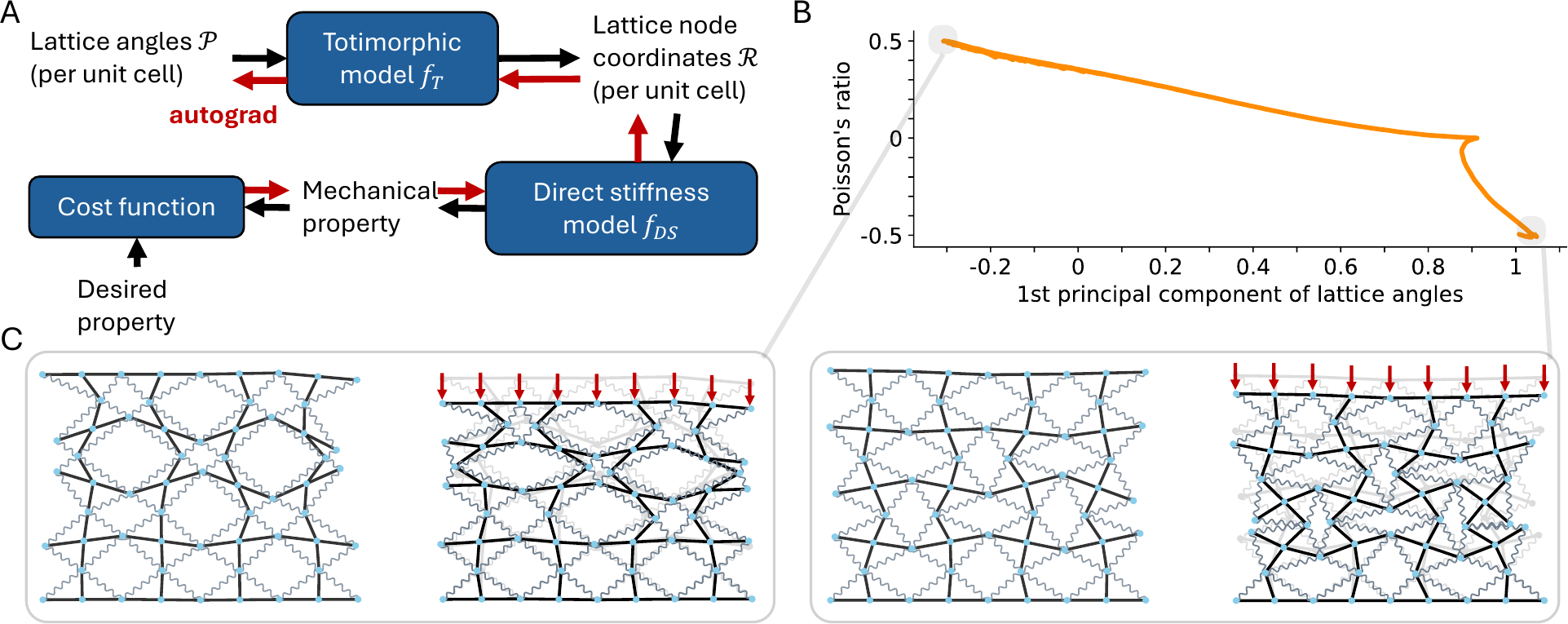}
	\caption{Inverse design of mechanical properties of Totimorphic lattices. \textbf{(A)} Schematic of the inverse design pipeline. Reconfiguration is driven by gradient descent on a cost function, realized using automatic differentiation (autograd). \textbf{(B)} During optimisation, we obtain a continuous trajectory through configuration space (here represented using the first principal component of the angles representing the lattice) connecting the initial configuration ($\nu = 0$) and the two final configurations ($\nu = \pm 0.5$). Since beams and levers are never deformed during reconfiguration, the lattice passes through all intermediate Poisson's ratios while morphing. \textbf{(C)} Illustration of the final design for $\nu = 0.5$ (left) and $\nu = - 0.5$ (right), shown without (left) and with load (right, red arrows).}	
	\label{fig:invDesign}
\end{figure*}

For the direct stiffness approach, we only consider beam and lever elements in the lattice, as they are the major supporting elements of the lattice.
We model these using Generalised Euler-Bernoulli beam elements \cite{ochsner2018finite}, described in \cref{methods:Euler}. 
Furthermore, we assume that joints in the Totimorphic lattice are locked during the compression experiment, meaning that the parameters $\totcoord$ are not changed while applying a load.
To do a single step of the compression experiment, we first calculate the stiffness matrix from the lattice points $\lattcoord$. 
Using the stiffness matrix, we can then derive the displacement of each node in the lattice resulting from the compression by solving a linear equation.
We repeat this process several times and note down the horizontal and vertical strain (mean length change).
Finally, the Poisson's ratio, which is defined as the slope of the strain-strain curve, is obtained via linear regression.
This process can be formalised as a single differentiable function $\femfunc$ which takes as input the physical lattice coordinates and returns the Poisson's ratio $\nu = \femfunc\left(\totfunc(\totcoord)\right)$.
The trajectory in parameter space ($\frac{\mathrm{d}}{\mathrm{d}t} \totcoord$) to reconfigure the lattice until it has a desired target Poisson's ratio $\nu^\mathrm{tgt}$ is then obtained by descending the gradient of the cost function $C$,
\begin{align}
 C(\totcoord) = &\| \nu^\mathrm{tgt} - \femfunc\left(\totfunc(\totcoord)\right) \|_1 \,, \\
\frac{\mathrm{d}}{\mathrm{d}t}& \totcoord \propto -\nabla_\totcoord C(\totcoord) \,,
\end{align}
where $\| \cdot \|_1$ is the L1 norm.
Further details about the mathematical formulation of the direct stiffness method and the Poisson's ratio can be found in \cref{methods:DS}.
The whole optimisation pipeline is illustrated in Fig.~\ref{fig:invDesign}A.

\subsubsection{Morphing between auxetic and non-auxetic configurations}
We demonstrate this approach for a $4\times4$ Totimorphic lattice with beam length $\beam = 1\,\mathrm{cm}$, beam cross-area $A=4\,\mathrm{mm}^2$, Young's modulus of the beam material of $E = 1\,\mathrm{GPa}$ (around the order of magnitude of polymers), and second moment of area $I= \frac{A^2}{12}$ (representing a square-shaped cross-sectional area).
The initial lattice has a Poisson's ratio of approximately zero, which is not surprising as it resembles a square lattice.
From this initial configuration, we perform gradient descent to get to a configuration with Poisson's ratio $\nu^\mathrm{tgt} = \pm 0.5$.
This results in two trajectories starting from the initial configuration and ending at a configuration with the desired Poisson's ratio, shown in Fig.~\ref{fig:invDesign}B using the first principal component of the lattice parameters $\mathcal{P}$.
The final lattice configurations and their response to applying load is shown in Fig.~\ref{fig:invDesign}C. 
In case of the configuration with positive Poisson's ratio (Fig.~\ref{fig:invDesign}C, left), large pockets were formed that widen during compression.
In case of negative Poisson's ratio (Fig.~\ref{fig:invDesign}C, right), the resulting lattice geometry is composed of square-shaped elements (one beam and two levers surrounded by springs) that pivot inwards when compressed from above.
This effect is most pronounced in the lower half of the lattice structure.
Similar designs were obtained for larger lattices, e.g., $8 \times 8$ trained to achieve even higher Poisson's ratios (Fig.~S1 in the Supplemental Information).
It is worthy noticing that every intermediate configuration obtained during optimisation is a legitimate Totimorphic lattice, and hence we are capable of continuously adjusting the Poisson's ratio by continuously changing the lattice angles.
For an animation of the reconfiguration process and the compression experiment (after reconfiguration), see Movie S1a,b and Movie S2a,b.
\review{Interestingly, the final configurations are not perfectly symmetric. First of all, symmetry is not required to minimise the cost function, so slight asymmetries are not too surprising. However, asymmetries are further caused by (i) the setup of the Totimorphic lattice, where for instance the bottom left node is fixed in space, while the top right node is strongly constrained by the remaining lattice, and (ii) limited numerical precision in the forward calculation of lattice points.}
Further simulation details are in \cref{sim:invLattice}.

\subsection{Second proof of concept: deployable and reconfigurable large-scale structures}\label{sec:totiscope}

As a second proof of concept, we present a telescope mirror with reconfigurable surface shape, evaluated using ray-based optics simulations. 
We envision that the Totimorphic lattice (on the length scale of meters) acts as a skeleton for, e.g., a graphene sheet that forms the reflective surface of a primary mirror of a space telescope \cite{rabien2023adaptive}\review{, or as a reconfigurable support truss for antennas \cite{tang2024deployable,wang2025shape}}.
We assume a $6\times6$ Totimorphic lattice with $\beam = 1$m in its flat surface configuration.

\subsubsection{Mirror deployment}
\begin{figure*}[t!]
    \centering
    \includegraphics[width=\textwidth]{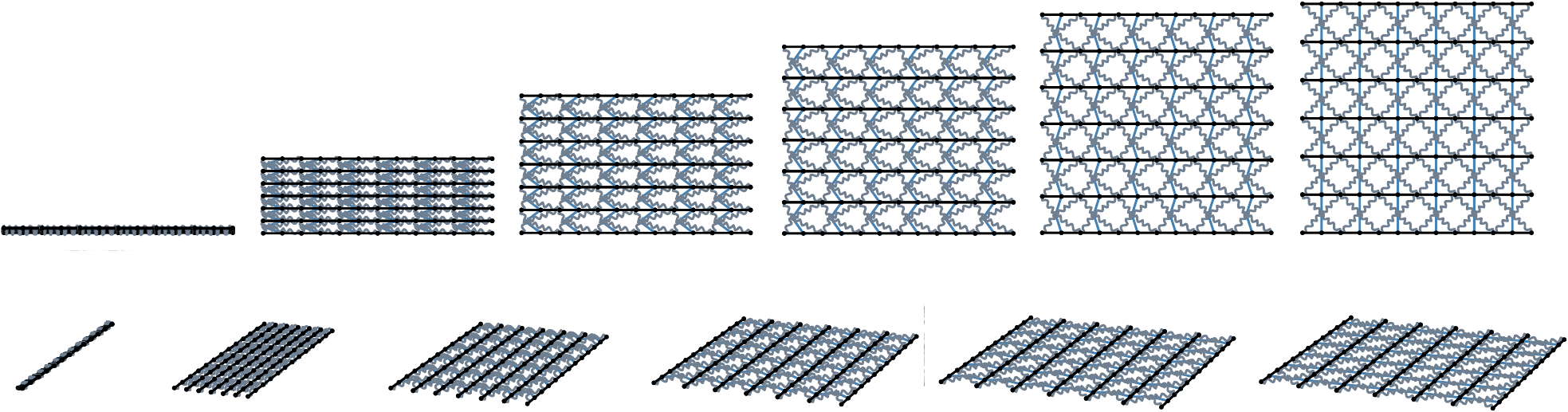}
	\caption{Deployment of a Totimorphic lattice structure. A Totimorphic sheet reconfigures from its collapsed (left) to its unfolded (right) configuration by raising all levers. The unfolding is shown from two different perspectives.}
	\label{fig:Unfold}
\end{figure*}

By only changing the lever angle of all elements in the lattice (i.e., setting them all to the same value), we can switch from the flat surface configuration to a collapsed one (and vice versa), greatly reducing the surface area and volume required to launch the Totimorphic structure (see Fig.~\ref{fig:Unfold}).
The reduction in surface area is given by $\sin\theta_\mathrm{min}$, where $\theta_\mathrm{min}$ is the minimum angle the levers can be set to physically.
Consequently, for launching, the lattice is configured in its collapsed configuration and for deployment, the lever angles are slowly increased to unfold the lattice into a flat surface.

\subsubsection{Continuously adjusting the mirror's focal point}
\begin{figure*}[t!]
    \centering
    \includegraphics[width=\textwidth]{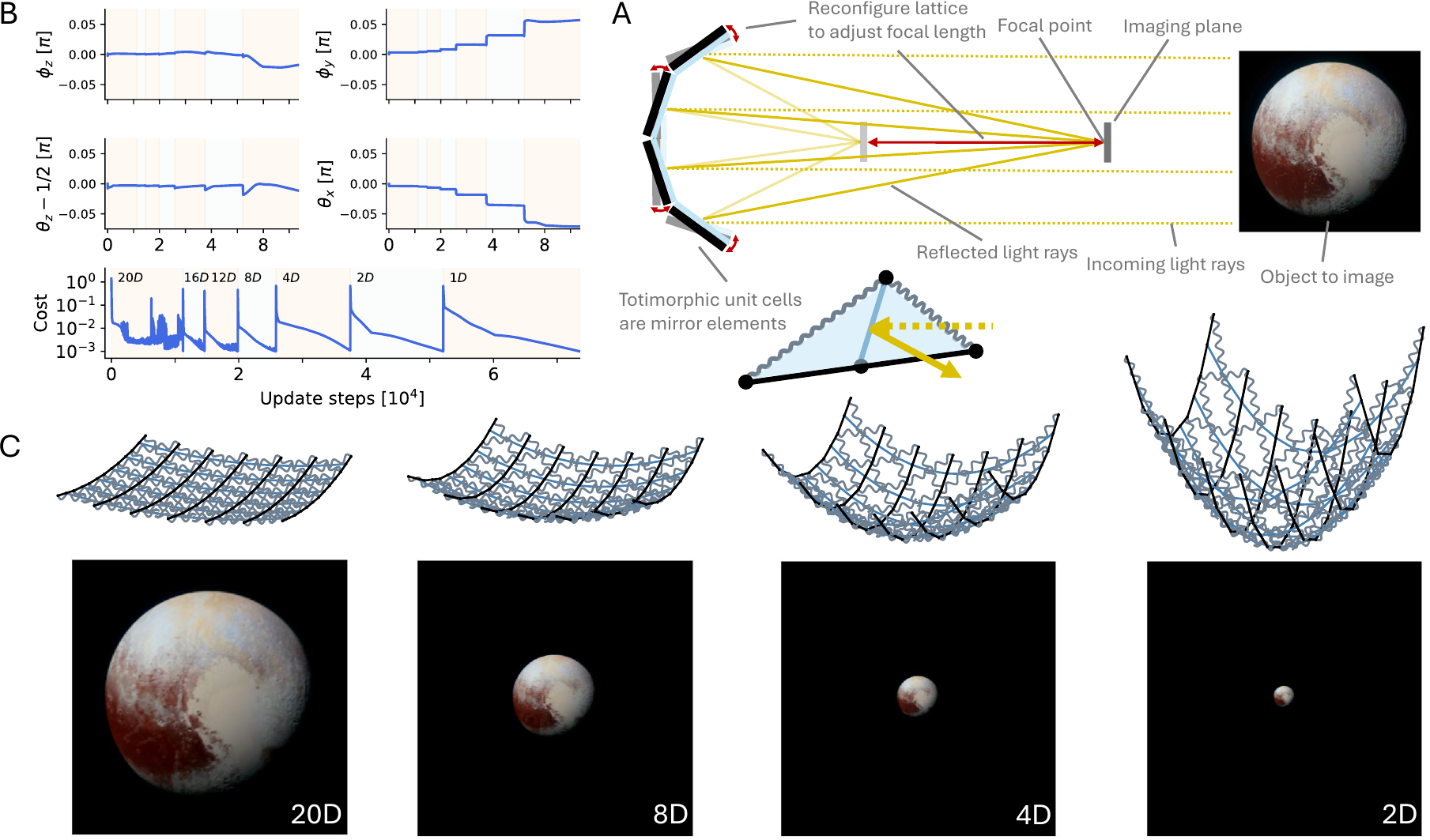}
	\caption{Deployment and reconfiguration of a Totimorphic telescope mirror. \textbf{(A)} Schematic of the studied setup. Given an object to be imaged, the unit cells of the Totimorphic lattice are used as mirror elements to focus the light on an imaging plane. By changing the lattice configuration, the focal point and focal length are adjusted.\textbf{(B)} Reconfiguration of the lattice to several different focal lengths, starting from a flat configuration. A few parameters as well as the objective-dependent cost are shown. \textbf{(C)} Lattice configuration and images received when using the lattice as a mirror, for configurations featuring different focal lengths. Credit for image of Pluto: NASA/JHUAPL/SwRI.}\vspace{-4mm}	
	\label{fig:telescopeDeployment}
\end{figure*}

After deployment, the lattice configuration is continuously adjustable to change the optical properties of its mirror surface.
In our case, the goal is to focus the light onto a secondary optical element such as a detector or a secondary mirror located at point $\focal$, which we call the target focal point in the following (Fig.~\ref{fig:telescopeDeployment}A).
We utilise a similar approach as for the material inverse design use-case, meaning that we construct a differentiable framework that allows us to find suitable lattice configurations (and trajectories connecting them) using gradient-based optimisation.
First, for simplicity, we assume that each unit cell reflects light at the mid-point of the lever in simulations (i.e., the mid-point between $\pC$ and $\pP$ as well $\pC$ and $\pPp$, see Fig.~\ref{fig:telescopeDeployment}A).
In addition, we assume that the object to be imaged is far away such that the incoming light is parallel.
Given an incidence direction $\pmb I$ ($\| \pmb I \|_2 = 1$, where $\| \cdot \|_2$ is the Euclidean norm) of the incoming light, the direction of the reflected light originating from each unit cell $(ij)$ is calculated using ray-based optics.
Subsequently, we obtain the point where each reflected light ray is closest to the target focal point (see \cref{methods:ray} for details).
The whole calculation can again be formulated as a series of differentiable operations, which we summarise in two functions: the Totimorphic model ($\totfunc$), and the optics part $\rayfunc$, such that  $\pmb F = \rayfunc(\totfunc(\totcoord))$ is a list of vectors containing the closest point to $\focal$ on each reflected light ray.

To find a Totimorphic configuration that focuses the light in a desired target focal point $\focal$, we again optimise the parameters characterising the Totimorphic lattice using gradient descent on a task-dependent cost function $C$,
\begin{align}
    C(\totcoord) = \sum_{\mathrm{unit\ cells}\ i} &\left\| \focal - \rayfunc\big(\totfunc(\totcoord)\big)_i  \right\|_2\,, \\
    \frac{\mathrm{d}}{\mathrm{d}t} \totcoord \propto& -\nabla_\totcoord C(\totcoord) \,,
\end{align}
During the optimisation process, we only present light coming from one direction (i.e., orthogonal to the initial lattice surface) which has to be focused into a single target point.
In Fig.~\ref{fig:telescopeDeployment}B-C, we show how the Totimorphic lattice transforms -- starting from a flat surface -- into curved surfaces that focus the light into a single point at decreasing focal lengths given in multiples of the mirror diameter $D$.
Fig.~\ref{fig:telescopeDeployment}B (bottom) shows the value of the cost function during deployment (from flat to a focal length of 20$D$, see Fig.~S2 in the Supplemental Information for further illustration) and during reconfiguration to new targets, i.e., focal lengths of 16, 12, 8, 4, 2, and 1$D$.
In all cases, we reach the same cost after reconfiguring the lattice, which means that the light reflected from all unit cells is focused with a similar precision in the target focal points.
In Fig.~\ref{fig:telescopeDeployment}B (top), a few lattice parameters are shown during the reconfiguration process, demonstrating that only minor continuous changes are required to achieve the desired result.
The reconfiguration is further illustrated in Movie S3a,b.

As a proof of concept, we use the Totimorphic lattice to image a real object; specifically, a photograph of Pluto at the same distance and angular extension of the Moon (as seen from Earth), see Fig.~\ref{fig:telescopeDeployment}A.
To construct the image, we record where light coming from each pixel of the image hits the imaging plane after being reflected by the unit cells of the lattice -- imitating the real process of image generation in a telescope (see \cref{methods:imaging} for details).
The resulting image closely resembles the original object, resolving small details such as contours of icy regions on Pluto (Fig.~\ref{fig:telescopeDeployment}C).
Moreover, as expected, decreasing the focal length of the Totimorphic mirror increases its field of view; enabling adaptive zoom-in and out by reconfiguring the lattice.

\begin{figure*}[t!]
    \centering
    \includegraphics[width=0.8\textwidth]{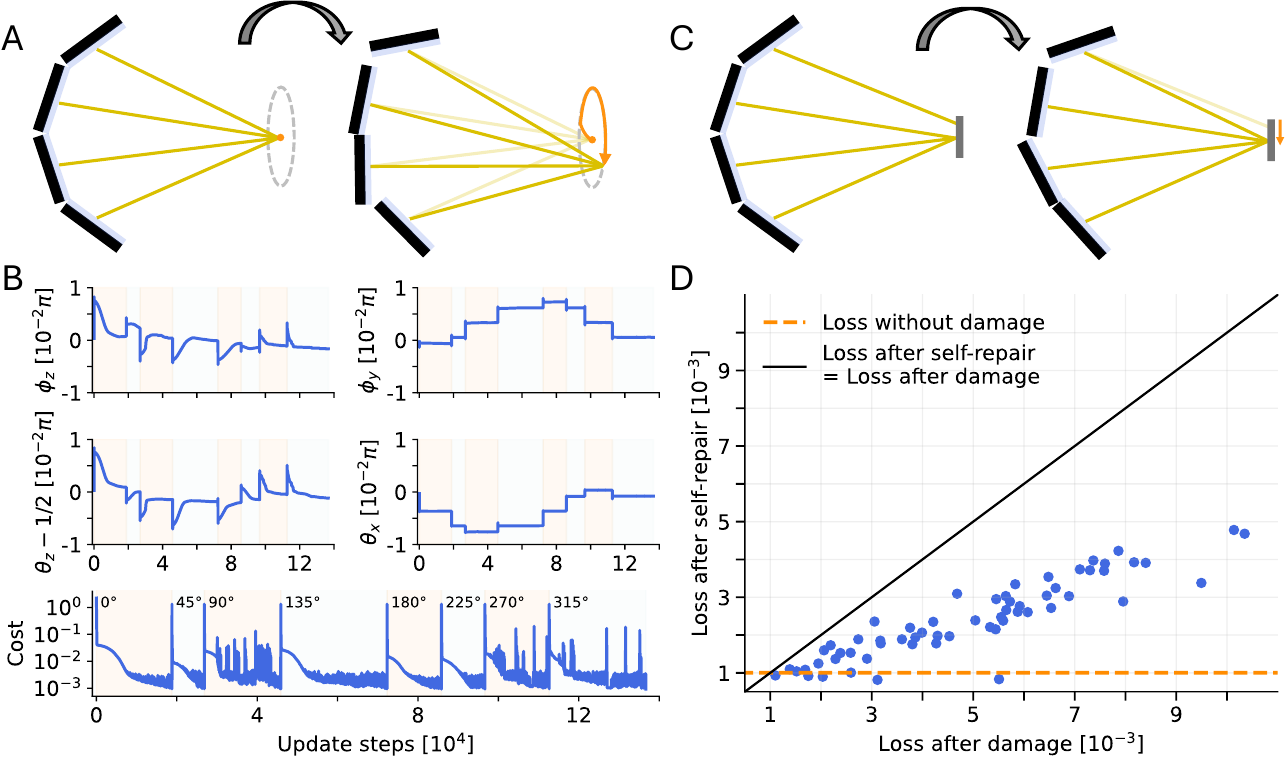}
	\caption{Off-center focus and self-repair capabilities. \textbf{(A)} The target focal point can be set off-center, which is shown here by guiding the focal point of the telescope on a circle around the original focal point. \textbf{(B)} Selected angle parameters of the lattice as well as the cost during the circular guidance experiment depicted in (A). During optimisation, several discrete target focal points at different angles on the circle were used. \textbf{(C)} Damage on one of the mirror elements (here: top) adds a deflection to light reflected by this element, thus putting it out of focus. Through reconfiguration of the lattice, the added deflection can be compensated for. \textbf{(D)} Result for self-repairing a single defect, repeated several times (blue dots). In each repetition, the damaged unit cell and the strength of the damage were selected randomly.}\vspace{-4mm}	
	\label{fig:telescopeAdaptation}
\end{figure*}

Although we only demonstrate reconfiguration for a few discrete focal lengths, in principle, the telescope can achieve a continuum of focal points by successively adjusting the target focal length.
In addition, the telescope is not limited to focal points that are on the central axis of the telescope surface, as shown in Fig.~\ref{fig:telescopeAdaptation}A-B where we guide the focal point on a circle around this axis using the above gradient-based optimisation.

\subsubsection{Self-repair}

The presented approach also allows to use reconfiguration of the lattice surface to compensate for defects introduced by mirror damage, e.g., through micrometeoroids \cite{rigby2023science}.
We illustrate this by adding a single defect to one of the lattice's unit cells (selected randomly), which is modelled by adding a deflection (randomly determined but static) to the light rays reflected from this cell (Fig.~\ref{fig:telescopeAdaptation}C; see \cref{methods:damage} for details).
As before, we perform gradient descent on the respective cost function to find a lattice configuration that compensates for this defect (Fig.~\ref{fig:telescopeAdaptation}C-D).
This is studied for increasing degrees of damage, i.e., for increasing amounts of deflection.
For all cases, the reconfiguration improves the performance of the mirror -- although for more extreme damage, we are not able to return to the baseline performance from before the damage occurs.
This is mainly because correcting for the damage requires an asymmetric configuration in the lattice, which is only possible to a limited degree before interfering with the focus capability of other (undamaged) unit cells.
But still, for smaller amounts of damage, reconfiguration allows a return to baseline performance.
In a realistic scenario, this could be realised in a ``lattice-in-the-loop'' setting, meaning that we use the actual lattice to collect where light from each unit cell hits the imaging plane, and then use the measurements in the backward pass (i.e., the computational graph used to calculate gradients for the parameters from the cost function) to get the required parameter changes of the physical mirror. Or put simply: using the simulation model as a proxy for the error calculation and replacing predicted values -- the distances to the focal point -- with real measurements.
\review{The same approach can be applied for other damage types as well, such as damaged actuators which would be modelled by keeping the corresponding generalised coordinate fixed or by adding strong penalty terms to the cost function.}
Further simulation details are in \cref{sim:telescope}.

\section{Discussion}

We introduced a differentiable parametrisation of Totimorphic lattices in two and three dimensions that enables constraint-free, gradient-based reconfiguration. Unlike most previously proposed reprogrammable systems \cite{lee2022mechanical,liu2022triclinic}, Totimorphic lattices can vary their effective properties continuously through geometry alone. 
Furthermore, by absorbing constraints into generalised coordinates, smooth trajectories can be generated using gradient descent instead of only final states. The method is simple to implement, compatible with modern automatic differentiation frameworks, and scalable to complex objectives, including cost functions that contain differentiable surrogate models for property prediction such as deep neural networks.

In our first application, we showed that a Totimorphic lattice can be reconfigured to continuously shift its Poisson’s ratio between negative and positive values.
This was illustrated for two-dimensional truss lattices. Extending this principle to three-dimensional materials would require designing volume-filling Totimorphic tilings and employing 3D stiffness methods, but the underlying approach remains the same.
In our second application, we explored Totimorphic lattices as the backbone for autonomously deployable and reconfigurable large-scale space infrastructure, focusing on a primary telescope mirror.
A key criterion for the operability of such a mirror is how accurately it focuses the reflected light, as this directly limits the attainable resolution when imaging objects.
For objects at far larger distances, e.g., with angular sizes similar to Jupiter, we can already see that the magnification of our design does not yet achieve sharp imaging -- although rough details are still recognisable (see Fig.~S3 in the Supplemental Information).
This can be improved by further increasing the focal length.
An approach that is often used in modern telescopes is to guide the light through a hierarchy of mirrors, e.g., by adding a secondary or even tertiary mirror before projecting the light onto a sensor.
In our case, those mirrors would be Totimorphic as well.
Due to the differentiability  of the whole process, our framework can be used to (re)configure all mirrors simultaneously.
This way, subsequent mirrors can even compensate for distortions introduced by previous (or later) mirrors.
In addition, the focusing of a mirror based on a Totimorphic lattice can be improved by reducing the size of unit cells. The effect of miniaturisation on focal length and the physical properties of the Totimorphic cells must be quantified to mature this concept.

Our current models also highlight methodological challenges. For the model in two dimensions, we formulated simple rules for detecting when lattice elements collide during reconfiguration.
In contrast, we did not employ any collision detection methods for the three-dimensional model, restricting our experiments to cases where strong folding of the lattice surface does not occur.
\review{In future work, approaches used to simulate textiles \cite{baraff2023large} could be employed to detect self-collisions in Totimorphic structures.
Apart from simply detecting whether the structure collides with itself, it is even more important to deal with self-collisions and invalid lattice configurations during the optimisation process.
A default method for dealing with collisions in, e.g., simulations of swarming robots \cite{izzo2007autonomous} or textile simulations \cite{baraff2023large} is to add strong repulsive forces that only act between elements at very low distances -- also known as barrier functions.
Adopting such an approach for Totimorphic structures is certainly feasible, but might be problematic due to the rigid and discontinuous nature of lattices. Furthermore, this method only helps in dealing with self-collisions and increases the computational cost of simulations substantially.
Alternatively, one could truncate parameter updates during optimisation such that general invalid lattice configurations are avoided (as proposed in \cite{fisherCloth2024}) -- although the optimisation process can get stuck trying to pass through invalid configurations, requiring either stochasticity to escape such local minima of the cost function or intermixing with optimisers that allow discontinuous changes of the lattice configuration.
In the presented experiments, we only checked whether the lattice is still valid after each update and stopped in case the lattice broke, which mostly occurred when attempting to reach too extreme mechanical properties.}

\review{Model-wise, instead of reparametrising the lattice, a variety of other techniques can be applied as well \cite{platt1987constrained}.
As an example, for gradient-based optimisation the \textit{penalty method} can be used where the beam and lever length and angle constraints are added as a penalty terms in the cost function to be reinforced.
However, it comes with the downside that the constraints are not perfectly satisfied during optimisation (hence we do not obtain trajectories), and in practice it can be non-trivial to avoid becoming stuck in local minima when trying to balance the penalty and the objective-dependent terms in the cost function.
Another alternative is the method of \textit{Lagrange multipliers}, but it does not work with gradient descent, requiring more involved solvers capable of finding saddle points of the cost function.
An extension of the Lagrange multiplier method, called \textit{basic differential multiplier method} \cite{platt1987constrained}, which is compatible with gradient descent can be used instead, although -- as in the case of the penalty method -- constraints are not satisfied throughout the whole optimisation process.
Apart from methods using gradient descent, more advanced approaches that divide the problem into simpler sub-problems can be used, such as sequential quadratic programming \cite{schittkowski1986nlpql}, although these are also not guaranteed to produce continuous trajectories.}

On the theory side, it is an open question whether there exists always a continuous trajectory connecting two valid Totimorphic configurations.
This is particularly important when utilising optimisation methods that -- different to gradient descent -- facilitate a global exploration of the configuration space, such as evolutionary strategies, to avoid selecting solutions that are not reachable from the initial lattice configuration. 
On the practical side, developing an actual prototype of a Totimorphic structure will require assessing the abstractions assumed in our model. For instance, in simulations we assume that the beams and levers are perfectly rigid. In reality, deformations in the structure from external or internal forces will produce bending in the beams, destabilising the internal spring forces. Similarly, the zero-length property of the springs will likely not hold for a real structure. \review{While it is possible to create springs that approximate the property for a range of lengths, at small or large extensions the zero-length property will likely not persist. Moreover, when in the zero-stiffness state, the Totimorphic structure should require no work to move between configurations, but practically there will be friction and jamming in the joints of the structure, which requires work from actuators to overcome. Independent of the application, minimising the number of actuators required to operate a spacecraft is quite critical, as each actuator represents a mode of failure that might compromise the spacecraft. Vacuum-exposed mechanisms and actuators are also at risk of cold welding and are difficult to lubricate. Since the number of actuators required to control a Totimorphic lattice scales linearly with the number of unit cells, it will be critical to investigate whether it is possible to reduce the number of actuators and optimise the operations required to morph between given configurations.The specifics of transforming 1D springs, Euler beams and frictionless joints into a manufacturable model is beyond the scope of this paper, but it will be necessary to make decisions that balance the requirements with suitability for space on various aspects like material choices, beam/lever aspect ratios, joint types, and actuation methods.
In future studies, the presented model should be extended to include such effects to reduce the gap between theory and practice.}

Despite these challenges, Totimorphic lattices have a number of advantages for use in the space environment. Deployable, reconfigurable, low-density structures are highly desirable for the strict mass and energy requirements of most space missions. We suggest that Totimorphic structures potentially fill a functional gap between \review{non-rigid deployables like fabrics and inflatables, and rigid materials like origami structures}, being able to carry tensile and compressive loads while also being flexible and capable of supporting surfaces without requiring pressurisation. \review{Unlike either inflatable or origami structures, Totimorphic lattices are highly reconfigurable and could fulfil many different functions during a single deployment.}
In particular, Totimorphic structures may be compatible with a variety of additional space infrastructure use-cases such as space antennas and solar sails.

To conclude, Totimorphic structures represent an intriguing class of mechanically reprogrammable systems, featuring continuously adjustable effective physical properties through simple mechanical and reversible actuations.
Especially since our framework is light-weight and flexible regarding the cost function, it can be implemented on edge devices to allow autonomous and lattice-aware control of the actuators, e.g., using a deep neural network to estimate effective properties of the Totimorphic structure from the state of its actuators (the generalised coordinates).
Thus, we are confident that the introduced concept can be scaled up to real-world, large-scale prototypes capable of autonomously and continuously adapting their lattice configuration to achieve a specified objective.\\

\section{Methods}\label{sec:methods}

In the following, we describe the mathematical models referenced in the main text, including mathematical proofs. 
These are just one particular realisation of an analytical Totimorphic model, and the introduced framework can be extended to other Totimorphic structures in a straightforward way.
Implementations of the models in Python are available on Github~\cite{github}. 
Simulation details are provided at the end of this section.

\subsection{Notation}
We first introduce the notation used in the following sections.
Polar vectors are denoted by:
\begin{equation}
\polar{\alpha} =   \begin{pmatrix} \cos{\alpha} \\ \sin{\alpha}
\end{pmatrix} \,.
\end{equation}
For proofs, we will be using the identity:
\begin{equation}
    \polar{\alpha} \cdot \polar{\beta} = \text{cos}\left(\alpha-\beta\right) \,. \label{eq:2DcosIdentity}
\end{equation}
We also introduce the  operator $\invPolar{\cdot}: \mathbb{R}^2 \to \mathbb{R}$ that returns the angles $\alpha$ required to represent a vector $\pmb r$ in polar coordinates, given by:
\begin{equation}
    \alpha = \invPolar{\pmb r} = \text{sign}\left(r_y \right) \cdot \arccos{\left(\frac{r_x}{\| \pmb r \|}\right)} \,.
\end{equation}
Rotation matrices in three-dimensional Euclidean space are given by:
\begin{equation}
\rot{x}{\alpha} = \begin{pmatrix}
1 & 0 & 0\\
0 & \cos{\alpha} & -\sin{\alpha} \\
0 & \sin{\alpha} & \cos{\alpha}
\end{pmatrix} \,,
\end{equation}  
\begin{equation}
\rot{y}{\alpha} = \begin{pmatrix}
\cos{\alpha} & 0 & \sin{\alpha}\\
0 & 1 & 0 \\
-\sin{\alpha} & 0 & \cos{\alpha}
\end{pmatrix} \,,
\end{equation}  
\begin{equation}
\rot{z}{\alpha} = \begin{pmatrix}
\cos{\alpha} & -\sin{\alpha} & 0\\
\sin{\alpha} & \cos{\alpha} & 0 \\
0 & 0 & 1
\end{pmatrix} \,.
\end{equation}  
Similarly, we denote the standard polar vector in three dimensions by:
\begin{equation}
\spherical{\alpha}{\beta} =   \begin{pmatrix} \cos{\alpha}\sin{\beta} \\ \sin{\alpha} \sin{\beta} \\ \cos{\beta}
\end{pmatrix} \,.
\end{equation}
To gain the angles for representing a general vector $\pmb r$ in these coordinates, we introduce $\invSph{\cdot}: \mathbb{R}^3 \to \mathbb{R}^2$ given by:
\begin{equation}
\begin{pmatrix} \alpha \\ \beta\end{pmatrix}
= \invSph{\pmb r}
= \begin{pmatrix} \text{sign}\left(r_y \right) \cdot \arccos{\left(\frac{r_x}{\sqrt{r_x^2 + r_y^2} }\right)} \\ \arccos\left(\frac{r_z}{\|\pmb r\|}\right)\end{pmatrix}
\end{equation}
We further use the following polar representation:
\begin{equation}
\sphericaly{\alpha}{\beta} =   \begin{pmatrix} \cos{\alpha}\cos{\beta} \\ \sin{\beta} \\ -\sin{\alpha}\cos{\beta}
\end{pmatrix} = \rot{y}{\alpha} \rot{z}{\beta} \pmb{1}_x
\end{equation}
with
\begin{equation}
\pmb{1}_x =   \begin{pmatrix} 1 \\ 0 \\ 0
\end{pmatrix}
\end{equation}
To gain the angles for representing a general vector $\pmb r$ in these coordinates, we introduce $\invSphy{\cdot}: \mathbb{R}^3 \to \mathbb{R}^2$ given by:
\begin{equation}
\begin{pmatrix} \alpha \\ \beta\end{pmatrix}
= \invSphy{\pmb r}
= \begin{pmatrix} -\text{sign}\left(r_z \right) \cdot \arccos{\left(\frac{r_x}{\sqrt{r_x^2 + r_z^2} }\right)} \\ \arcsin\left(\frac{r_y}{\|\pmb r\|}\right)\end{pmatrix}
\end{equation}
For proofs, the following identity is used:
\begin{equation}
     \spherical{\alpha}{\beta} \cdot \spherical{\phi}{\Theta} = \cos\left(\alpha - \phi \right) \sin{\beta} \sin{\Theta} + \cos{\beta}\cos{\Theta} \,. \label{eq:sphIdentity}
\end{equation}

\subsection{Totimorphic lattices in two dimensions}\label{sec:methodslattice2D}

\subsubsection{Parametrisation of elementary cells.}

In the following, we drop the index $ij$ (cell at row $i$ and column $j$) for clarity. However, all vectors are always given per elementary cell (see main text). An elementary cell is constructed from two unit cells which are connected at their levers.
An elementary cell in the Totimorphic lattice is parameterised by $(\origin{}, \beamAngle{}, \leverAngle{}, \beamAnglep{}, \leverAnglep{})$, where $\origin{}$ is the location of point $A$, $\beamAngle{}$ is the angle with the x-axis of the lower beam, and $\leverAngle{}$ is the angle of the lower lever with respect to the lower beam.
Primed variables represent the same quantities, but for the upper half of the elementary cell (i.e., the upper beam and lever).

\noindent The points of the lower unit cell are given by:
%
    \begin{align}
        \rvec{}{\pA} &= \origin{} \,, \\
        \rvec{}{\pB} &= \rvec{}{\pA} + \beam  \cdot \polar{\beamAngle{}} \,, \\
        \rvec{}{\pP} &= \rvec{}{\pA} + \lever \cdot \polar{\beamAngle{}} \,, \\
        \rvec{}{\pC} &= \rvec{}{\pP} + \lever \cdot \polar{\beamAngle{}+\leverAngle{}} \,.
    \end{align}
%
The points of the upper unit cell are given by:
%
    \begin{align}
        \rvec{}{\pPp} = \rvec{}{\pC} + \lever \polar{\invBeamAnglep{}}&\ \ \ \ \text{with} \ \ \ \ \invBeamAnglep{} = \pi - \leverAnglep{} + \beamAnglep{} \,, \\
        \rvec{}{\pAp} &= \rvec{}{\pPp} - \lever \cdot \polar{\beamAnglep{}} \,, \\
        \rvec{}{\pBp} = \rvec{}{\pPp} + &\lever \cdot \polar{\beamAnglep{}} = 2 \cdot \rvec{}{\pPp} - \rvec{}{\pAp} \,.
    \end{align}
%
For the upper triangle, we recover $\beamAnglep{}$ via
%
    \begin{align}
        \polar{\beamAnglep{}} &= \frac{1}{\beam} \cdot \left(\rvec{}{\pBp} - \rvec{}{\pAp}\right) \,, \\ 
        \beamAnglep{} &= \text{sign}\left(\polary{\beamAnglep{}}\right) \cdot \text{arccos}\left(\polarx{\beamAnglep{}}\right) \,, \label{eq:phi}
    \end{align}
%
and $\leverAnglep{}$ through
%
    \begin{align}
        \polar{\invBeamAnglep{}} &= \frac{2}{\beam} \cdot \left(\rvec{}{\pPp} - \rvec{}{\pC}\right) \,, \\ 
        \invBeamAnglep{} &= \text{sign}\left(\polary{\invBeamAnglep{}}\right) \cdot \text{arccos}\left(\polarx{\invBeamAnglep{}}\right) \,, \\
        \leverAnglep{} &= \pi - \invBeamAnglep{} + \beamAnglep{} \,,
    \end{align}
%
with $\beamAnglep{}$ given by Eq.~\ref{eq:phi}.

To check for overlapping springs in the elementary cell, representing a state in which the cell physically breaks, we introduce the following angles per elementary cell:
%
    \begin{align}
        \stopleft{} &= \frac{\leverAnglep{}}{2} - \beamAnglep{} \,, \\
        \stopright{} &= \frac{\pi}{2} - \stopleft{} \,, \\
        \sbotleft{} &= \frac{\leverAngle{}}{2} + \beamAngle{} \,, \\
        \sbotright{} &= \frac{\pi}{2} - \sbotleft{} \,.
    \end{align}
%
$\sbotleft{}$ is the angle of the bottom left spring (connecting points $A$ and $C$) with the x-axis, while $\sbotright{}$ is the angle of the bottom right spring (connecting points $B$ and $C$).
Primed angles indicate the same for the top unit cell.
Angles are measured with respect to the x-axis going through point $C$, hence in the flat configuration, top angles are negative and bottom angles are positive.

\subsubsection{Connecting elementary cells in two dimensions}\label{sec:methodsQuadratic}

The first elementary cell ($i=0$, $j=0$) can be constructed freely, as there are no other cells to connect it to.
Adding more unit cells horizontally ($i=0$, $j > 0$) is done iteratively as follows.
First, we enforce the constraint of connecting the two elementary cells:
%
    \begin{align}
 \rvec{0j}{\pA} &= \rvec{0j-1}{\pB} \,, \\
 \rvec{0j}{\pAp} &= \rvec{0j-1}{\pBp} \,.
    \end{align}
%
Then we have to obtain the point $P'$ of our new elementary cell, which has to satisfy the beam and lever constraints.
In the following, all points are in the same elementary cell $(0j)$, and thus we drop the indices for convenience. 
The condition for the existence of $\rvec{}{\pPp}$ is that lever and beam with fixed length have to connect to it from point $A'$ (given by the first elementary cell) and point $C$: 
%
\begin{align}
\|\rvec{}{\pAp} - \rvec{}{\pPp}\|^2 &= \left(\lever\right)^2 \,, \\
\|\rvec{}{\pC} - \rvec{}{\pPp}\|^2  &=  \left(\lever\right)^2 \,.
\end{align}
%
Introducing $\springm{} = \rvec{}{\pAp} - \rvec{}{\pC}$, $\springp{} = \rvec{}{\pAp} + \rvec{}{\pC}$ and $\delta_{} = \frac{1}{2} \left( \| \rvec{}{\pAp} \|^2 - \| \rvec{}{\pC} \|^2 \right)$, we get

\begin{align}
    \rvecxn{\pPp} &= \frac{\delta_{} - \rvecyn{\pPp} \springmyn}{\springmxn} \,, \label{eq:2dPx}\\
    \rvecyn{\pPp} &= \frac{1}{2} \left(\springpyn - \springmxn \sqrt{ \left(\frac{\beam}{\| \springm{}\|} \right)^2 - 1} \right) \,.\label{eq:2dPy}
\end{align}

\subsubsection{Proof: derivation of expression for $\pPp$}\label{si:p2D}

In the following, we do not write the indices $ij$ explicitly to simplify the notation.

First, we start from the condition that $\pPp$ has to be chosen such that the resulting beam (connecting $\pAp$ and $\pPp$) and lever (connecting $\pC$ and $\pPp$) have the desired length:
%
\begin{align}
\|\rvec{}{\pAp} - \rvec{}{\pPp}\|^2 &= \left(\lever\right)^2 \,,  \label{SI:eq1} \\
\|\rvec{}{\pC} - \rvec{}{\pPp}\|^2  &=  \left(\lever\right)^2 \,. \label{SI:eq2}
\end{align}
%
Expanding both expressions, subtracting Eq.~\ref{SI:eq2} from Eq.~\ref{SI:eq1} and dividing the resulting equation by $2$, we obtain:
\begin{equation}
\frac{1}{2}\left(\|\rvec{}{\pAp}\|^2 - \|\rvec{}{\pC}\|^2\right) - \rvecyn{\pPp} \left(\rvecyn{\pAp} - \rvecyn{\pC}\right) - \rvecxn{\pPp} \left(\rvecxn{\pAp} - \rvecxn{\pC}\right) = 0 \,.
\end{equation}
Using the notation $\springm{} = \rvec{}{\pAp} - \rvec{}{\pC}$, $\springp{} = \rvec{}{\pAp} + \rvec{}{\pC}$ and $\delta_{} = \frac{1}{2} \left( \| \rvec{}{\pAp} \|^2 - \| \rvec{}{\pC} \|^2 \right)$ as in the main text, this yields Eq.~\ref{eq:2dPx}.
To obtain Eq.~\ref{eq:2dPy}, we first insert Eq.~\ref{eq:2dPx} into Eq.~\ref{SI:eq2} and rewrite it to match the form of a quadratic equation:
\begin{align}
    &(\rvecyn{\pPp})^2 + p \cdot \rvecyn{\pPp} + q = 0 \,, \label{si:pqform} \\
    \mathrm{with} \ \ \ & p = \frac{2 \rvecxn{\pC} \springmyn \springmxn - 2 \rvecyn{\pC} (\springmxn)^2 - 2 \delta \springmyn}{(\springmxn)^2 + (\springmyn)^2} \,, \\
    & q = \frac{\delta^2 - 2 \delta \rvecxn{\pC} \springmxn + (\rvecxn{\pC} \springmxn)^2 + (\rvecyn{\pC} \springmxn)^2 - (\springmxn)^2 (\lever)^2}{(\springmxn)^2 + (\springmyn)^2} \,.
\end{align}
Writing out all terms in the numerator of $p$ (denoted by $p_\mathrm{n}$), it can be simplified to
\begin{align}
    p_\mathrm{n} = \ & 2 \rvecxn{\pC} \springmyn \springmxn - 2 \rvecyn{\pC} (\springmxn)^2 - 2 \delta \springmyn \nonumber \\ 
    = \ & 2 \rvecxn{\pC} \rvecxn{\pAp} \rvecyn{\pAp} - (\rvecxn{\pC})^2 \rvecyn{\pAp} - \rvecyn{\pC} (\rvecxn{\pAp})^2 + 2 \rvecxn{\pC} \rvecyn{\pC} \rvecxn{\pAp} - (\rvecxn{\pAp})^2 \rvecyn{\pAp} \nonumber \\
    & - (\rvecyn{\pAp})^3 + (\rvecyn{\pC})^2 \rvecyn{\pAp} + (\rvecyn{\pAp})^2 \rvecyn{\pC} - (\rvecxn{\pC})^2 \rvecyn{\pC} - (\rvecyn{\pC})^3 \,.
\end{align}
Using the two identities:
\begin{align}
    &(\rvecyn{\pAp})^3 - (\rvecyn{\pAp})^2\rvecyn{\pC} + \rvecyn{\pAp} (\rvecxn{\pC})^2 - \rvecyn{\pAp} (\rvecyn{\pC})^2 + (\rvecxn{\pC})^2 \rvecyn{\pC} + (\rvecyn{\pC})^3 = (\rvecyn{\pAp} + \rvecyn{\pC})\left[(\rvecyn{\pAp} - \rvecyn{\pC})^2 + (\rvecxn{\pC})^2 \right] \,,\\
    & \rvecyn{\pC} (\rvecxn{\pAp})^2 - 2 \rvecxn{\pAp} \rvecyn{\pAp} \rvecxn{\pC} - 2 \rvecxn{\pC} \rvecyn{\pC} \rvecxn{\pAp} + (\rvecxn{\pAp})^2 \rvecyn{\pAp} = (\rvecyn{\pAp} + \rvecyn{\pC})\left[(\rvecxn{\pAp} - \rvecxn{\pC})^2 - (\rvecxn{\pC})^2 \right] \,,
\end{align}
we arrive at
\begin{equation}
    p_\mathrm{n} = - \springpyn \left[(\springmxn)^2 + (\springmyn)^2 \right] \,,
\end{equation}
and thus
\begin{equation}
    p = - \springpyn \,.
\end{equation}
For $q$, we first note (using binomial formulas) that 
\begin{equation}
    \delta - \rvecxn{\pC} \springmxn = \frac{1}{2} \left[(\rvecxn{\pAp} - \rvecxn{\pC})^2 + (\rvecyn{\pAp}-\rvecyn{\pC})(\rvecyn{\pAp}+\rvecyn{\pC}) \right] \,.
\end{equation}
Using $\delta^2 - 2 \delta \rvecxn{\pC} \springmxn + (\rvecxn{\pC} \springmxn)^2 = (\delta - \rvecxn{\pC} \springmxn)^2$ as well as $(\frac{p}{2})^2 = \frac{1}{4} (\rvecyn{\pAp} + \rvecyn{\pC})^2 \left[(\rvecxn{\pAp} - \rvecxn{\pC})^2 + (\rvecyn{\pAp} - \rvecyn{\pC})^2\right]$, we can then calculate the remaining term $(\frac{p}{2})^2 - q$ required for solving Eq.~\ref{si:pqform}. First. we only look at all terms in $(\frac{p}{2})^2 - q$ not containing $\beam$. The numerator of this term is given by:
\begin{align}
   &- \frac{1}{4} \left[(\rvecxn{\pAp} - \rvecxn{\pC})^2 + (\rvecyn{\pAp}-\rvecyn{\pC})(\rvecyn{\pAp}+\rvecyn{\pC}) \right]^2 - (\rvecyn{\pC})^2 (\rvecxn{\pAp} - \rvecxn{\pC})^2 + \frac{1}{4} (\rvecyn{\pAp} + \rvecyn{\pC})^2 \left[(\rvecxn{\pAp} - \rvecxn{\pC})^2 + (\rvecyn{\pAp} - \rvecyn{\pC})^2\right] \,, \nonumber \\
   = \ & - \frac{1}{4} (\springmxn)^2 \left[ (\rvecxn{\pAp} - \rvecxn{\pC})^2 + 2 ( (\rvecyn{\pAp})^2 - (\rvecyn{\pC})^2 ) - (\rvecyn{\pAp} + \rvecyn{\pC})^2 + 4 (\rvecyn{\pC})^2 \right] \,, \\
   = \ & - \frac{1}{4} (\springmxn)^2 \left[(\springmxn)^2 + (\springmyn)^2 \right] \,.
\end{align}
The terms containing $\beam$ simplify to $\frac{\beam^2 (\springmxn)^2}{4 \| \springm{} \|^2}$. Using the previous results, we arrive at:
\begin{equation}
    \left(\frac{p}{2}\right)^2 - q = \frac{1}{4} (\springmxn)^2 \cdot \left(\left(\frac{\beam}{\| \springm{} \|}\right)^2 - 1 \right)\,.
\end{equation}
Finally, putting everything together, we can solve $Eq.~\ref{si:pqform}$ using the p-q formula and by inputting the results for $p$ and $\left(\frac{p}{2}\right)^2 - q$:
\begin{equation}
    \rvecyn{\pPp} = \frac{1}{2} \left(\springpyn \pm \springmxn \sqrt{ \left(\frac{\beam}{\| \springm{}\|} \right)^2 - 1} \right) \,.
\end{equation}
$\frac{1}{2} \springp{}$ is the mid-point of the spring connecting point $C$ and $A'$. 
Hence, $\springpyn$ is its $y$ coordinate (or height).
If point $A'$ is left of point $C$, as e.g. in the initial flat configuration, point $P'$ has to be above this spring; meaning that its $y$ coordinate has to be larger than $\springpyn$.
Since $\springmxn$ is negative in this scenario, the minus sign has to be chosen to get the correct solution.
Similarly, if we pivot the unit cell such that $\springmxn = 0$, i.e., $A'$ and $C$ lie above each other, we have that $\rvecyn{\pPp} = \frac{1}{2} \springpyn$.
Pivoting further, $P'$ is situated below the spring (although it never swapped sides), as given by Eq.~\ref{eq:2dPy}.
Thus, when starting from the flat hour-glass motive, we have to choose the negative sign to get the correct solution for $P'$.

\subsubsection{Conditions for intact elementary cells in two dimensions}\label{sec:methodsAngles}

The criteria for being intact are the same for every unit cell $(ij)$, therefore indices are dropped again in the following to ease notation.
An elementary cell is considered intact if the following conditions are met:
\begin{align}
    \| \springm{} \ \| &\leq \beam \,, \label{eq:2Dcond}\\
    \stopleft{} + \sbotleft{} &\geq 0 \,, \\
    \stopright{} + \sbotright{} &\geq 0 \,.
\end{align}
The first condition means that springs cannot be stretched further than the beam length (for a solution of $\rvec{}{\pPp}$ to exist, Eq.~\ref{eq:2dPy} has to be a real number, meaning that the argument of the square root has to be larger or equal to zero. This immediately yields Eq.~\ref{eq:2Dcond}).
The second (third) conditions check whether the left (right) bottom and top springs cross.

To avoid overstretched springs, we correct $\leverAngle{}$ by constraining its range. The limits of this interval are obtained by the criterion for the existence of real-valued solutions of Eq.~\ref{eq:2dPy} (the term below the root has to be $\geq 0$):
\begin{equation}
    \Delta_{} - \arccos{\left(  C \right)} \leq \leverAngle{} \leq 
    \Delta_{} + \arccos{\left(  C \right)} \,. \label{eq:2dThetaRange}
\end{equation}
with $\weirdvec{} = \rvec{}{\pAp} - \rvec{}{\pA} - \lever \polar{\beamAngle{}}$, $\Phi_{} = \invPolar{\weirdvec{}}$, $\Delta_{} = \Phi_{} - \beamAngle{}$, and $C_{} = \frac{\| \weirdvec{} \|}{\beam} - \frac{3}{4} \frac{\beam}{\| \weirdvec{} \|}$. 
If the interval for $\leverAngle{}$ is empty, it is impossible to add a new elementary cell without changing the configuration of previous cells in the structure.
These conditions are checked for each unit cell separately while constructing it.

\subsubsection{Proof: deriving allowed range of $\leverAngle{}$}\label{sec:thetaRange}

First, we write out $\rvec{}{\pC}$ in $\springm{}$:
%
\begin{align}
    \springm{} &= \rvec{}{\pAp} - \rvec{}{\pC} \,,\\
    &= \rvec{}{\pAp} - \rvec{}{\pA} - \lever \cdot \polar{\beamAngle{}} - \lever \cdot \polar{\beamAngle{}+\leverAngle{}} \,,\\
    &= \weirdvec{} - \lever \cdot \polar{\beamAngle{}+\leverAngle{}}\,,
\end{align}
%
with $\weirdvec{} = \rvec{}{\pAp} - \rvec{}{\pA} - \lever \polar{\beamAngle{}}$.
This way, we split up $\springm{}$ into two terms -- one depending on $\leverAngle{}$, and one that is independent of $\leverAngle{}$, summarising the geometric constraints introduced by the previous unit cell.
By representing $\weirdvec{}$ in polar coordinates, we arrive at
\begin{equation}
    \springm{} = \|\weirdvec{}\| \polar{\Phi} - \lever \cdot \polar{\beamAngle{}+\leverAngle{}}\,, \label{eq:SI2dspring}
\end{equation}
with $\Phi_{} = \invPolar{\weirdvec{}}$, which simplifies the following derivation of the range of $\leverAngle{}$.
To guarantee that a solution for point $\pPp$ exists, Eq.~\ref{eq:2Dcond} has to be fulfilled. 
By squaring Eq.~\ref{eq:2Dcond} and using Eq.~\ref{eq:SI2dspring}, we get
%
\begin{align}
 \left\| \|\weirdvec{}\| \polar{\Phi} - \lever \cdot \polar{\beamAngle{}+\leverAngle{}} \right\|^2  \leq \beam^2 \hspace{4mm} &\Leftrightarrow \hspace{4mm} \|\weirdvec{}\|^2 - \beam \|\weirdvec{}\|  \polar{\Phi} \cdot \polar{\beamAngle{}+\leverAngle{}} + \frac{1}{4}\beam^2 \leq \beam^2 \,, \\
 &\Leftrightarrow \hspace{4mm} \polar{\Phi} \cdot \polar{\beamAngle{}+\leverAngle{}} \geq \frac{\| \weirdvec{} \|}{\beam} -\frac{3}{4} \frac{\beam}{\| \weirdvec{} \|} \,, \\
  &\Leftrightarrow \hspace{4mm} \cos\left(\Phi - \beamAngle{} - \leverAngle{}\right) \geq \frac{\| \weirdvec{} \|}{\beam} -\frac{3}{4} \frac{\beam}{\| \weirdvec{} \|} \,, \\
  &\Leftrightarrow \hspace{4mm} \cos\left(\Delta - \leverAngle{}\right) \geq C \,,
\end{align}    
%
where we used Eq.~\ref{eq:2DcosIdentity} and introduced $\Delta_{} = \Phi_{} - \beamAngle{}$, and $C_{} = \frac{\| \weirdvec{} \|}{\beam} - \frac{3}{4} \frac{\beam}{\| \weirdvec{} \|}$.
From this, we can invert the cosine function to get
\begin{equation}
    \pm \left(\Delta - \leverAngle{}\right) \leq \arccos{C} \,,
\end{equation}
from which we get the upper (choosing $-$) and lower (choosing $+$) bound of $\leverAngle{}$ shown in Eq.~\ref{eq:2dThetaRange}.

\subsubsection{Building a two-dimensional Totimorphic structure}

For $i > 0$ and $j=0$, one sets $\beamAngle{i0} = \beamAnglep{i-1,0}$, $\rvec{i,0}{\pA} = \rvec{i-1,0}{\pAp}$, and calculates all points as for the first elementary cell ($i=0$, $j=0$).
In all remaining cases, we set $\beamAngle{ij} = \beamAnglep{i-1,j}$, $\rvec{i,j}{\pA} = \rvec{i,j-1}{\pB}$, $\rvec{ij}{\pAp} = \rvec{i,j-1}{\pBp}$, and calculate the remaining points by solving for $\pPp$.

\noindent Depending on where in the lattice an elementary cell is positioned, it is characterised by the following parameters:
\begin{enumerate}
    \item $i=0$, $j=0$: $(\origin{00}, \beamAngle{00}, \leverAngle{00}, \beamAnglep{00}, \leverAnglep{00})$ \,,
    \item $i=0$, $j>0$: $(\beamAngle{0j}, \leverAngle{0j})$ \,,
    \item $i>0$, $j=0$: $(\leverAngle{i0}, \beamAnglep{i0}, \leverAnglep{i0})$ \,,
    \item else: $(\leverAngle{ij})$ \,.
\end{enumerate}
To initialise the lattice as a flat surface, beam angles $\beamAngle{ij}$ and $\beamAnglep{ij}$ have to be set to $0$ and lever angles $\leverAngle{ij}$ and $\leverAnglep{ij}$ to $\frac{\pi}{2}$ $\forall i,j$.

\subsection{Totimorphic lattices in three dimensions}\label{sec:methodslattice3D}

\subsubsection{Parametrisation of elementary cells.}

As in the two-dimensional case, each elementary cell is characterised by the angles of the beams and levers, $(\origin{}, \beamAngley{}, \beamAnglez{}, \leverAnglex{}, \leverAnglez{}, \beamAngleyp{}, \beamAnglezp{}, \leverAnglexp{}, \leverAnglezp{})$.

\noindent The points of the lower unit cell are given by:
%
    \begin{align}
        \rvec{}{\pA} &= \origin{} \,, \\
        \rvec{}{\pB} &= \rvec{}{\pA} + \beam  \cdot \sphericaly{\beamAngley}{\beamAnglez} \,, \\
        \rvec{}{\pP} &= \rvec{}{\pA} + \lever \cdot \sphericaly{\beamAngley}{\beamAnglez} \,, \\
        \rvec{}{\pC} &= \rvec{}{\pP} + \lever \cdot \rot{y}{\beamAngley} \rot{z}{\beamAnglez} \rot{x}{\leverAnglex} \rot{z}{\leverAnglez} \pmb{1}_x\,.
    \end{align}
%
Here, the beam is described by its rotation in the x-y plane (given by $\beamAnglez{}$, which corresponds to $\beamAngle{}$ in the two-dimensional case) and a rotation in x-z plane (a tilting of the unit cell given by $\beamAngley$).
The orientation of the lever is obtained by a sequence of rotations: first, we rotate the x-direction unit vector $\pmb{1}_x$ by $\leverAnglez$ (in the two-dimensional case, we called this $\leverAngle{}$), then tilt it in the y-z plane by $\leverAnglex$, and then apply the same transformation (two rotations) used to gain the beam vector from the unit vector.
We choose this parametrisation so that the angles of the lever are again defined with respect to the orientation of the beam, as in the two-dimensional case.

\noindent The points of the upper unit cell are defined similarly:
%
    \begin{align}
        \rvec{}{\pPp} &= \rvec{}{\pC} + \lever \cdot \rot{y}{\beamAngleyp} \rot{z}{\beamAnglezp} \rot{x}{-\leverAnglexp} \rot{z}{-\leverAnglezp} \pmb{1}_x \,, \\
        \rvec{}{\pAp} &= \rvec{}{\pPp} - \lever \cdot \sphericaly{\beamAngleyp}{\beamAnglezp} \,, \\
        \rvec{}{\pBp} &= \rvec{}{\pPp} + \lever \cdot \sphericaly{\beamAngleyp}{\beamAnglezp} \,.
    \end{align}
%

\subsubsection{Connecting elementary cells in three dimensions}

Again, the first elementary cell ($i=0$, $j=0$) is unconstrained and can be constructed freely.
Without loss of generality, we initialise lattices lying in the x-z plane.
Adding more unit cells horizontally ($i=0$, $j > 0$) is done iteratively as follows.
First, we enforce the constraint of connecting
the two elementary cells:
%
    \begin{align}
 \rvec{0j}{\pA} &= \rvec{0j-1}{\pB} \,, \\
 \rvec{0j}{\pAp} &= \rvec{0j-1}{\pBp} \,.
    \end{align}
%
In the following, all points are in the same elementary cell $(0j)$, and thus we drop the indices for convenience. 
As in the two-dimensional case, we look again for a consistent solution for $\rvec{}{\pPp}$.
The vector representing the upper beam is given by $\beam \cdot \spherical{\phi}{\Theta}$ (the traditional angles are recovered via $(\beamAngley{}, \beamAnglez{}) = \invSphy{\spherical{\phi}{\Theta}}$.
The lever is then given by the vector
%
\begin{align}
\pmb r_L &= \rvec{}{\pAp} - \rvec{}{\pC} + \lever \spherical{\phi}{\Theta}  \,, \\
&= \springm{} + \lever \cdot \spherical{\phi}{\Theta} \,.
\end{align}
%
For $P'$ to exist, the lever has to have the correct length: 
\begin{equation}
\| \pmb r_L \|^2 = \lever^2  \,. \label{eq:beamSol}
\end{equation}
From this, we obtain a solution for $\phi$ and $\Theta$, with $\phi$ being completely determined by $\Theta$.
Additionally, as in the two-dimensional case, we gain valid intervals for both $\Theta$ as well as $\leverAnglez$.
The remaining points are then given by:
%
    \begin{align}
        \rvec{}{\pPp} = \rvec{}{\pAp} + \lever \cdot \spherical{\phi}{\Theta} \,, \\
        \rvec{}{\pBp} = \rvec{}{\pAp} + \beam  \cdot \spherical{\phi}{\Theta}  \,.
    \end{align}
%

\subsubsection{Proof: deriving expressions for $\phi$ and $\Theta$}

Expanding Eq.~\ref{eq:beamSol}, we get
\begin{equation}
    \springm{} \cdot \spherical{\phi}{\Theta} = - \frac{\| \springm{} \|^2}{\beam} \,.
\end{equation}
Writing $\springm{}$ in spherical coordinates with $(\alpha, \beta) = \invSph{\springm{}}$, this becomes
\begin{equation}
    \spherical{\alpha}{\beta} \cdot \spherical{\phi}{\Theta} = - \frac{\| \springm{} \|}{\beam} \,,
\end{equation}
or, using Eq.~\ref{eq:sphIdentity}, 
\begin{equation}
\spherical{\alpha}{\beta} \cdot \spherical{\phi}{\Theta} = \cos\left(\alpha - \phi \right) \sin{\beta} \sin{\Theta} + \cos{\beta}\cos{\Theta} = - \frac{\| \springm{} \|}{\beam} \,,    
\end{equation}
from which we get the solution for $\phi$ depending on $\Theta$:
\begin{align}
\phi = \alpha - \arccos\left(\mathcal{C}\right) \hspace{4mm} \text{with} \hspace{4mm} \mathcal{C} = - \frac{\frac{\| \springm{} \|}{\beam} + \cos{\beta}\cos{\Theta}}{ \sin{\beta} \sin{\Theta}} \,.
\end{align}
The arccos can have two signs, so as in the two-dimensional case, two solutions exist. We found that when starting with the flat surface configuration, a minus sign has to be chosen. However, different from the two-dimensional case, the alternative solution ($+$ instead of $-$) also represents a valid Totimorphic configuration, although the two obtained configurations are far apart.
In the case $\sin{\beta} = 0$, $\phi$ can be freely chosen and $\Theta$ has a unique solution $\Theta = \arccos\left(- \frac{\| \springm{} \|}{\beam \cos{\beta}} \right)$.
In this case, the  spring vector $\springm{}$ is parallel to the z-axis. Any solution given by $\Theta$ will now form a triangle, from $\pAp$ to $\pPp$, from $\pPp$ to $\pC$, and from $\pC$ back to $\pAp$ (which is parallel to the z-axis). Changing $\phi$ will now rotate this triangle around the z-axis without distorting it, hence there is an infinite amount of valid configurations that lie on a circle around the z-axis.

If $\sin{\beta} \neq 0$ (i.e., $\beta \neq 0$, which is almost always the case as $\beta = 0$ is a singular point), $\mathcal{C}$ has to be in the allowed range of $\arccos$:
\begin{equation}
    -1 \leq \mathcal{C} \leq 1 \,. \label{eq:siCond3d}
\end{equation}
Note that in spherical coordinates, $\beta \in [0, \pi]$, and hence we always have $\sin\beta \geq 0$ (and similarly $\sin\Theta \geq 0$).
From $\mathcal{C} \geq -1$, we get 
\begin{align}
    - \frac{\| \springm{} \|}{\beam} \geq \cos{\beta}\cos{\Theta} -\sin{\beta}\sin{\Theta}\,.
\end{align}
and, by using $\sin\left(2\pi-\beta\right) = -\sin{\beta}$, $\cos\left(2\pi-\beta\right) = \cos{\beta}$, and $\cos\left(2\pi - \beta\right)\cos{\Theta}+\sin\left(2\pi - \beta\right)\sin{\Theta} = \cos\left(2\pi - \beta - \Theta\right)$,
\begin{align}
    - \frac{\| \springm{} \|}{\beam} \geq \cos\left(2\pi - \beta - \Theta\right)\,.
\end{align}
which yields
\begin{align}
    \arccos\left( - \frac{\| \springm{} \|}{\beam}\right) - \beta \leq \Theta \leq 2\pi - \beta - \arccos\left(-\frac{\| \springm{} \|}{\beam}\right) \,. \label{eq:Theta1}
\end{align}
This range is valid when $\mathcal{C} \leq 0$.
If $\mathcal{C} \geq 0$, we have to guarantee that $\mathcal{C} \leq 1$, which results in
\begin{equation}
    \beta - \arccos\left( - \frac{\| \springm{} \|}{\beam}\right) \leq \Theta \leq \beta + \arccos\left( - \frac{\| \springm{} \|}{\beam}\right) \,. \label{eq:Theta2}
\end{equation}

\subsubsection{Proof: deriving allowed range of $\leverAnglez$}

Finally, we can calculate an allowed range for $\leverAnglez$ (as in the 2D case).
This can be derived by requiring that the $\arccos$ in Eq.~\ref{eq:Theta1} and~\ref{eq:Theta2} has a valid solution, i.e.,
\begin{equation}
    \| \springm{} \|^2 \leq \beam^2 \,. \label{eq:SIthetaCond}
\end{equation}
First, we rewrite $\springm{}$ by expanding it:
\begin{equation}
    \springm{} = \rvec{}{\pAp} - \rvec{}{\pC} = \rvec{}{\pAp} - \rvec{}{\pP} - \lever \cdot \rot{y}{\beamAngley} \rot{z}{\beamAnglez} \rot{x}{\leverAnglex} \rot{z}{\leverAnglez} \mathbb{1}_x
\end{equation}
To isolate $\leverAnglez$, we rotate $\springm{}$ by $\rot{x}{-\leverAnglex}\rot{z}{-\beamAnglez}\rot{y}{-\beamAngley}$, which gives us
\begin{equation}
    \rot{x}{-\leverAnglex}\rot{z}{-\beamAnglez}\rot{y}{-\beamAngley} \springm{} = \weirdvec{} - \lever \cdot \spherical{\leverAnglez}{\frac{\pi}{2}} \,,
\end{equation}
with $\weirdvec{} = \rot{x}{-\leverAnglex}\rot{z}{-\beamAnglez}\rot{y}{-\beamAngley}\cdot\left(\rvec{}{\pAp} - \rvec{}{\pP}\right)$. Since rotations leave the length of a vector unchanged, we can thus rewrite condition Eq.~\ref{eq:SIthetaCond} as
\begin{equation}
    \| \weirdvec{} - \lever \cdot \spherical{\leverAnglez}{\frac{\pi}{2}} \|^2 \leq \beam^2 \,.
\end{equation}
We solve this the same way as in Eq.~\ref{eq:2dThetaRange}, resulting in 
\begin{align}
\spherical{\theta_k}{\phi_k} \cdot \spherical{\leverAnglez}{\frac{\pi}{2}} \geq \frac{\| \weirdvec{} \|}{\beam} - \frac{3}{4} \frac{\beam}{\| \weirdvec{} \|} \,,      
\end{align}
with $(\theta_k, \phi_k) = \invSph{\weirdvec{}}$. 
Using $\spherical{\theta_k}{\phi_k} \cdot \spherical{\leverAnglez}{\frac{\pi}{2}} = \sin{\phi_k} \cdot \cos\left(\theta_k - \leverAnglez\right)$, we get
\begin{equation}
  \cos\left(\theta_k - \leverAnglez\right) \geq C \hspace{4mm} \text{with} \hspace{4mm} C = \frac{\| \weirdvec{} \|}{\beam \sin{\phi_k}} - \frac{3}{4} \frac{\beam}{\| \weirdvec{} \|  \sin{\phi_k}}  \,,
\end{equation}
from which the final interval is derived
\begin{equation}
   \theta_k - \arccos{C} \leq \leverAnglez \leq \theta_k + \arccos{C}\,.
\end{equation}

\subsubsection{Building a three-dimensional Totimorphic structure}

For $i > 0$ and $j=0$, one sets $\beamAnglezt{i0} = \beamAnglezpt{i-1,0}$, $\beamAngleyt{i0} = \beamAngleypt{i-1,0}$, $\rvec{i,0}{\pA} = \rvec{i-1,0}{\pAp}$, and calculates all points as for the first unit cell ($i=0$, $j=0$).
In all remaining cases, we set $\beamAnglezt{i0} = \beamAnglezpt{i-1,0}$, $\beamAngleyt{i0} = \beamAngleypt{i-1,0}$, $\rvec{i,j}{\pA} = \rvec{i,j-1}{\pB}$, $\rvec{ij}{\pAp} = \rvec{i,j-1}{\pBp}$, and calculate the remaining points by solving for $\phi$ and $\Theta$.

\noindent Depending on where in the lattice an elementary cell is positioned, it is characterised by the following parameters (we drop indices here for clarity, but all parameters are different):
\begin{enumerate}
    \item $i=0$, $j=0$: $(\origin{}, \beamAngley{}, \beamAnglez{}, \leverAnglex{}, \leverAnglez{}, \beamAngleyp{}, \beamAnglezp{}, \leverAnglexp{}, \leverAnglezp{})$
    \item $i=0$, $j>0$: $(\beamAngley{}, \beamAnglez{}, \leverAnglex{}, \leverAnglez{}, \Theta)$
    \item $i>0$, $j=0$:  $(\leverAnglex{}, \leverAnglez{}, \beamAngleyp{}, \beamAnglezp{}, \leverAnglexp{}, \leverAnglezp{})$
    \item else: $(\leverAnglex{}, \leverAnglez{}, \Theta)$
\end{enumerate}
To initialise the lattice as a flat surface in x-z plane, beam angles $\beamAnglezt{ij}$ and $\beamAnglezpt{ij}$ have to be set to $\frac{\pi}{2}$ and all remaining angles to 0.

\subsection{Methodology for the first proof of concept: direct stiffness method}\label{methods:DS}

The fundamental building block of the direct stiffness method is rod or beam-like elements characterised by the coordinates of its two ending points $i,j \in \mathbb{N}$, i.e., $(x_i, y_i)$ and $(x_j, y_j)$.
Applying forces $(\pmb F_i, \pmb F_j)$ to the two ending points of the element results in a deformation of the element; or more specifically, a displacement of the ending points $(\pmb u_i, \pmb u_j)$, which is calculated from the stiffness equation (a generalisation of Hook's law):
\begin{equation}
    \pmb{K}_{ij} \begin{pmatrix}
    \pmb u_i \\
    \pmb u_j
    \end{pmatrix} = 
    \begin{pmatrix}
    \pmb F_i \\
    \pmb F_j
    \end{pmatrix} \,, \label{eq:stiffness}
\end{equation}
with $\pmb{K}_{ij}$ being the stiffness matrix, and $\pmb u_i = (\xd{i}, \yd{i}, \phid{i})$ being the resulting displacements due to external forces and moments $\pmb F_i = (\Fx{i}, \Fy{i}, \Mphi{i})$. $\varphi$ characterises the resulting bending of beam elements.
Thus, the updated equilibrium state of the beam after applying forces is given by $(x_i + \xd{i}, y_i+\yd{i})$ and $(x_j+\xd{j}, y_j+\yd{j})$.
Repeating this process by recalculating the stiffness matrix using the updated coordinates and solving Eq.~\ref{eq:stiffness} anew, deformations due to larger forces can be obtained.

\subsubsection{Generalised Euler-Bernoulli beam elements}\label{methods:Euler}

In this work, we use generalised Euler-Bernoulli beam elements to model the beams and levers in a Totimorphic lattice.
The stiffness matrix $\pmb{K}_{ij}$ of the generalised beam element is obtained by combining the stiffness matrices of rod elements, i.e., elements that behave like springs,
\begin{equation}\label{eq:rod}
    \pmb{K}^\mathrm{rod}_{ij} = \frac{E_{ij} A_{ij}}{L_{ij}}
\begin{pmatrix}
c_{ij}^2 & c_{ij} s_{ij} & 0 & -c_{ij}^2 & -c_{ij} s_{ij} & 0 \\
c_{ij} s_{ij} & s_{ij}^2 & 0 &  -c_{ij} s_{ij} & -s_{ij}^2 & 0 \\
0 & 0 & 0 & 0 & 0 & 0 \\
-c_{ij}^2 & -c_{ij} s_{ij} & 0 & c_{ij}^2 & c_{ij} s_{ij} & 0 \\
-c_{ij} s_{ij} & -s_{ij}^2 & 0 & c_{ij} s_{ij} & s_{ij}^2 & 0 \\
0 & 0 & 0 & 0 & 0 & 0
\end{pmatrix} \,.
\end{equation}
and Euler-Bernoulli beam elements that model bending \cite{ochsner2018finite}
\begin{align}
    &\pmb{K}^\mathrm{EB}_{ij} = \frac{E_{ij} I_{ij}}{L_{ij}^3} \cdot \label{eq:beam}
    \begin{pmatrix}
    12 s_{ij}^2 & -12 s_{ij} c_{ij} & -6 L_{ij} s_{ij} & -12 s_{ij}^2 & 12 s_{ij} c_{ij} & -6 L_{ij} s_{ij} \\
    -12 s_{ij} c_{ij} & 12 c_{ij}^2 & 6 L_{ij} c_{ij} & 12 s_{ij} c_{ij} & - 12 c_{ij}^2 & 6 L_{ij} c_{ij} \\
    -6 L_{ij} s_{ij} & 6 L_{ij} c_{ij}  & 4 L_{ij}^2 & 6 L_{ij} s_{ij} & -6 L_{ij} c_{ij} & 2 L_{ij}^2 \\
    -12 s_{ij}^2 &  12 s_{ij} c_{ij} & 6 L_{ij} s_{ij} & 12 s_{ij}^2 & -12 s_{ij} c_{ij} & 6 L_{ij} s_{ij} \\
    12 s_{ij} c_{ij} & -12 c_{ij}^2 & -6 L_{ij} c_{ij} & -12 s_{ij} c_{ij} & 12 c_{ij}^2 & -6 L_{ij} c_{ij} \\
    -6 L_{ij} s_{ij} &  6 L_{ij} c_{ij}  & 2 L_{ij}^2 & 6 L_{ij} s_{ij} &  -6 L_{ij} c_{ij} & 4 L_{ij}^2
    \end{pmatrix} \,,
\end{align}
resulting in 
\begin{equation}
    \pmb{K}_{ij} = \pmb{K}^\mathrm{rod}_{ij} + \pmb{K}^\mathrm{EB}_{ij} \,.
\end{equation}
Here, $L_{ij}$ is the length of the beam, and $s_{ij}$ and $c_{ij}$ are the sine and cosine of the angle $\vartheta_{ij}$ of the beam with respect to the x-axis:
\begin{align}
    L_{ij} &= \sqrt{(x_j - x_i)^2 + (y_j - y_i)^2}\,, \\
    c_{ij} &= \cos\left(\vartheta_{ij}\right) \,, \\
    s_{ij} &= \sin\left(\vartheta_{ij}\right) \,.
\end{align}
$E_{ij}$, $A_{ij}$, and $I_{ij}$ are the elements Young's modulus, cross-sectional area, and second moment of area, respectively.
For simplicity, we choose a uniform lattice with $E_{ij} = E$, $I_{ij} = I$, and $A_{ij} = A$.

\subsubsection{Compression experiment}\label{methods:compression}

The global stiffness matrix $\pmb G$ of a whole lattice structure is obtained by adding up the stiffness matrices of the individual beams and levers accordingly.
The stiffness equation for the whole structure then takes the same form as the one for individual elements, but it contains all lattice point coordinates (ending points of Totimorphic beams or levers, i.e., points $A$, $A'$, $B$, $B'$, $P$, $P'$, and $C$ of each elementary cell):
\begin{equation}\label{eq:stiffnessEq}
     \pmb{G} \begin{pmatrix}
    \pmb u_0 \\
    \pmb u_1 \\
    ... \\
    \pmb u_{N-1}
    \end{pmatrix} = 
    \begin{pmatrix}
    \pmb F_0 \\
    \pmb F_1 \\
    ... \\
    \pmb F_{N-1}
    \end{pmatrix} \,.
\end{equation}\\
To implement the compression experiment, we first set the displacement of points that are part of the top surface of the lattice to a small non-zero value.
As boundary conditions, we set the displacement of all points in the bottom surface to zero (otherwise, the lattice would start moving downwards).
Using Eq.~\ref{eq:stiffnessEq} (i.e. matrix multiplication), we calculate the resulting forces for all other points due to these forced displacements -- from which we subsequently calculate the resulting displacements of all remaining points in the lattice that are not part of the top or bottom surface by solving a system of linear equations.
We repeat this process several times (with updated global stiffness matrix $\pmb G$) to obtain macroscopic deformations.
For a detailed description, see \cite{dold2023differentiable}.

\subsubsection{Poisson's ratio}\label{methods:Poisson}

Assume we do $k$ iterations in the compression experiment, i.e., we solve Eq.~\ref{eq:stiffnessEq} $k$ times in sequence, and in each iteration we displace the top lattice points by $\epsilon_k$.
After each iteration, we collect the displacement in $x$-direction of the outer left and outer right points of the lattice , calculate the mean displacement over these points, and then obtain the average change in width of the lattice by taking the difference:
\begin{align}
    \Delta R_k &= \frac{1}{|\mathcal{RS}|} \sum_{i \in \mathcal{RS}} \xd{i,k} \,, \\
    \Delta L_k &= \frac{1}{|\mathcal{LS}|} \sum_{i \in \mathcal{LS}} \xd{i,k} \,, \\
    \bar{\epsilon}_k &= \Delta R_k - \Delta L_k \,,
\end{align}
where $\mathcal{LS}$ ($\mathcal{RS}$) is a set containing the indices of the points on the left (right) surface of the lattice.
The outer points are given\textbf{ (left side)} by points $\pA$ and $\pAp$ of the first (i.e. outer left) column of elementary cells in the lattice, and \textbf{(right side)} by the points $\pB$ and $\pBp$ of the last column (i.e. outer right) of elementary cells.
$|\mathcal{RS}|$ denotes the number of elements in $\mathcal{RS}$.
The Poisson's ratio is obtained using linear regression
\begin{equation}
    \nu = \frac{\sum_k \bar{\epsilon}_k}{\sum_k \epsilon_k} \,.
\end{equation}
The deformation in vertical direction is imposed on the top points, i.e., $\pAp$ and $\pBp$ of all elementary cells in the top roow, while the bottom points ($\pA$ and $\pB$ of all elementary cells in the bottom row) are kept static.

\review{Alternative ways of calculating the Poisson ratio are possible, e.g., by imposing target displacements for the outer nodes in the cost function (e.g., no displacement to reach a Poisson ratio of 0, as done in \cite{dold2023differentiable}, Fig. 6C). Furthermore, to enforce symmetric behaviour during compression, additional terms can be added to the cost function that punish asymmetric deformations under load.}

\subsection{Methodology for the second proof of concept: ray-based optics}\label{methods:ray}

In the telescope mirror application, for simplicity, we only simulate light rays getting reflected at the mid-point (i.e. the mid-point of the lever) of each unit cell.
Thus, each unit cell acts as a tiny mirror reflecting a single beam of light.
We assume light coming from a source that is far away, such that the light rays hitting the different mirror elements are parallel to each other.

\subsubsection{Reflections}\label{methods:refl}

To calculate the light reflection, we have to know the orientation of the surface of reflection, which is given by the normal vector of each unit cell.
For a single elementary cell, the normal vectors of its two unit cells are
\begin{align}
    \nvec{ij} &= \frac{\left(\rvec{ij}{\pB} - \rvec{ij}{\pA}\right) \times \left(\rvec{ij}{\pC} - \rvec{ij}{\pA} \right)}{\left\| \left(\rvec{ij}{\pB} - \rvec{ij}{\pA}\right) \times \left(\rvec{ij}{\pC} - \rvec{ij}{\pA} \right) \right\|_2} \,, \\
    \nvecp{ij} &= \frac{\left(\rvec{ij}{\pBp} - \rvec{ij}{\pC}\right) \times \left(\rvec{ij}{\pPp} - \rvec{ij}{\pC} \right)}{\left\| \left(\rvec{ij}{\pBp} - \rvec{ij}{\pC}\right) \times \left(\rvec{ij}{\pPp} - \rvec{ij}{\pC} \right) \right\|_2} \,,
\end{align}
where $\|\cdot \|_2$ is the Euclidean norm, $\times$ the vector cross product, and the primed vector denotes the top unit cell.
The mid-points of the levers where the light is reflected are
\begin{align}
    \mvec{ij} &= \frac{1}{2} \left(\rvec{ij}{\pP} + \rvec{ij}{\pC} \right) \,,\\
    \mvecp{ij} &= \frac{1}{2} \left(\rvec{ij}{\pPp} + \rvec{ij}{\pC} \right) \,.
\end{align}
Given a light ray with incidence direction $\pmb I$ ($\| \pmb I\|_2 = 1$), the directions of the reflected light $\pmb l_{ij}$ coming from elementary cell $(ij)$ are given by
\begin{align}
    \pmb l_{ij} &= \frac{\pmb I - 2 \left(\pmb I \cdot \nvec{ij} \right) \nvec{ij}}{\| \pmb I - 2 \left(\pmb I \cdot \nvec{ij} \right) \nvec{ij}\|_2} \,, \\
    \pmb l'_{ij} &= \frac{\pmb I - 2 \left(\pmb I \cdot \nvecp{ij} \right) \nvecp{ij}}{\| \pmb I - 2 \left(\pmb I \cdot \nvecp{ij} \right) \nvecp{ij}\|_2} \,,
\end{align}
where $\cdot$ is the vector dot product.

\subsubsection{Smallest distance to the target focal point}\label{methods:closest}

Given a target focal point $\focal$, we calculate the closest point $\pmb c_{ij}$ ($\pmb c'_{ij}$) along the reflected ray to it as follows:
\begin{align}
    \pmb c_{ij} &= \mvec{ij} + \big((\focal - \mvec{ij}) \cdot \pmb l_{ij}\big) \pmb l_{ij} \,, \\
    \pmb c'_{ij} &= \mvecp{ij} + \big((\focal - \mvecp{ij}) \cdot \pmb l'_{ij}\big) \pmb l'_{ij} \,.
\end{align}
We then calculate the following cost function
\begin{equation}
    C(\totcoord, \focal) = \frac{1}{2N\cdot M} \sum_i^N \sum_j^M \big( \|\pmb c_{ij} - \focal \|^2 + \|\pmb c'_{ij} - \focal \|^2 \big) \,,
\end{equation}
and optimise the Totimorphic lattice by updating its parameters via gradient descent, $-\nabla_\totcoord C(\totcoord, \focal)$.

\subsubsection{Damaging mirror elements}\label{methods:damage}

To emulate a damaged mirror element, we add a random perturbation to the direction of the reflected light.
For instance, assuming the bottom unit cell of element $(ij)$ was damaged, we first sample a random deflection $\pmb l_\sigma \sim \mathcal{N}(0,\sigma^2)$ from a Gaussian distribution (with mean $0$ and variance $\sigma^2$) and always add it to the reflected light direction:
\begin{equation}
    \pmb l_{ij} \mapsto \frac{\pmb l_{ij} + \pmb l_\sigma}{\| \pmb l_{ij} + \pmb l_\sigma\|_2} \,.
\end{equation}
For a single experiment, $\pmb l_\sigma$ is only sampled once and kept constant afterwards.

\subsubsection{Imaging}\label{methods:imaging}

For imaging, we assume a picture of Pluto with the same diameter and distance as the Moon (radius of $r_m \approx 1700 \cdot 10^3\,\mathrm{m}$ and angular radius of $\phi_m \approx 0.25^\circ$) has been placed at the same distance as the Moon (given by $d_m = r_m / \arctan\phi_m$).

For a $2S \times 2S$ pixel image of Pluto, the light ray coming from the pixel in row $i$ and column $j$ (counted from the bottom left corner of the image) has the non-normalised incidence vector
\begin{equation}
    \tilde{\pmb{I}}_{ij} = \begin{pmatrix}
    \left(\frac{i}{S} - 1\right) r_m \\
    \left(\frac{j}{S} - 1\right) r_m \\
    - d_m
    \end{pmatrix} \,,
\end{equation}
and the normalised one $\pmb{I}_{ij} = \frac{\tilde{\pmb{I}}_{ij}}{\|\tilde{\pmb{I}}_{ij}\|_2}$.
To get an image, we reflect the light rays coming from every pixel in the image at every Totimorphic mirror element and check where on the imaging plane the reflections hit.
For instance, if the imaging plane is at distance $d_z$, the light ray $\pmb I_{kl}$ reflected at the bottom unit cell of element $(ij)$ hits the plane at $\mvec{ij} + \lambda \cdot \pmb l_{ij}$ with $\lambda = \left(d_z - m_{ij,z}\right)/l_{ij,z}$ ($z$ denoting the $z$-component of the vectors).
For visualisation, we plot all reflected points in a two-dimensional image with point size $s = 2$ (in matplotlib) and an alpha (opacity) of $0.05$, so that colours merge when multiple points overlap.

\subsection{Simulation details: inverse lattice design}\label{sim:invLattice}

For the results shown in Fig.~\ref{fig:invDesign}, during inverse design, we used a total external displacement of $0.2$ applied through $5$ iterations of the compression experiment.
The cost function was optimised using the Adam optimiser (learning rate of $5\cdot10^{-3}$ and no weight decay).
To ensure that the top layer of nodes in the lattice remains flat, we added regularisation terms to the cost function, i.e., L1 loss terms that encode the difference in y-coordinate for points in the top layer.
The training setup for negative and positive target Poisson's ratio was identical apart from initialisation.
For the positive target, we perturbed the initial configuration before training by adding Gaussian noise to each lattice parameter; drawn randomly from $\mathcal{N}\left(\mu = 0, \sigma = 10^{-2}\right)$.
To visualise the final lattice designs in Fig.~\ref{fig:invDesign} and Fig.~S1, we used a total external displacement of $0.325$ and $0.5$, respectively, applied over $30$ iterations instead.
For the results shown in Fig.~S1, we trained with a target Poisson's ratio of $\nu^\mathrm{tgt} = -2$.

\subsection{Simulation details: Totimorphic telescope}\label{sim:telescope}

We trained the lattice only on the task of focusing light from one direction (i.e., orthogonal to the initially flat mirror surface) into a single point (focal point on the central axis orthogonal to the initially flat mirror surface) located at $(\frac{D}{2}, \frac{D}{2}, n\cdot D$, where $D = 6$m is the diameter (or side length) of the Totimorphic lattice and $n$ an integer that we choose during training (e.g., we start with $20$).
For training configurations at different focal lengths, we used the Adam optimiser with a learning rate of $10^{-3}$, which is reduced to $10^{-4}$ if the root mean square loss goes below $5\cdot 10^{-3}$. 
We further always add a second cost term (with prefactor of $0.1$) which has the objective of keeping the central node (point $\pBp$ of the central elementary cell) of the lattice fixed in space.
This is done to avoid that the lattice moves around in space during optimisation.

For the circle experiment in Fig.~\ref{fig:telescopeAdaptation}, we guide the focal point in a circle with radius $\frac{D}{2}$ around its initial location (still with a focal length of $20 D$.
We use the following learning rate scheme for this: starting with a learning rate of $10^{-3}$, we reduce it to $10^{-4}$ and $10^{-5}$ when the root mean square loss goes below $5\cdot 10^{-2}$ and $5\cdot 10^{-3}$, respectively.

For the self-repair experiments in Fig.~\ref{fig:telescopeAdaptation}, we select the damaged unit cell randomly and draw the deflection due to damaging randomly from $\mathcal{N}\left(0, 5\cdot 10^{-4}\right)$. During training, we start with a learning rate of $10^{-4}$, which is reduced to $10^{-5}$ of the root mean square loss goes below $5\cdot 10^{-3}$.

\section*{Acknowledgments}
We would like to thank Derek Aranguren van Egmond, Michael Mallon, and Max Bannach for helpful and inspiring discussions, and our colleagues at ESA’s Advanced Concepts Team for their ongoing support. We further thank Leon Williams for helpful feedback on the manuscript.
AT, NR, and DD acknowledge support through ESA’s young graduate trainee and fellowship programs. DD further acknowledges support through Horizon Europe's Marie Sklodowska-Curie Actions (Project 101103062 — BASE).
\paragraph*{Author contributions:}
Initial idea is by AT and DD.
Theory and model implementation is by DD, based on initial work by AT.
DD, AT, NR, JG, and DI contributed to the design of the study and the performed experiments.
DD and NR implemented code for the proof of concept experiments.
DD wrote the first draft of the paper.
DD, AT, NR, JG, and DI read and reviewed the final paper.
\paragraph*{Competing interests:}
There are no competing interests to declare.
\paragraph*{Data and code availability:}
Implementations of the models in Python are available on GitHub \cite{github}.

\printbibliography
\addcontentsline{toc}{section}{References}

\clearpage
\renewcommand{\appendixname}{Supplemental Information}
\begin{center}
\Large{\textbf{Supplemental Information}}\\[10pt]
\end{center}
\renewcommand{\thesection}{\arabic{section}}  
\renewcommand{\thefigure}{S\arabic{figure}}
\renewcommand{\theequation}{S\arabic{equation}}
\setcounter{equation}{0}
\setcounter{figure}{0}
\setcounter{section}{0}
\setcounter{subsection}{0}
\setcounter{page}{1}
\renewcommand{\thetable}{S\arabic{table}}
\renewcommand{\thesection}{S\arabic{section}}
\setcounter{table}{0}
\addtocontents{toc}{\protect\setcounter{tocdepth}{0}}

\section*{Supplemental figures}

\begin{figure*}[ht!]
    \centering
    \includegraphics[width=172mm]{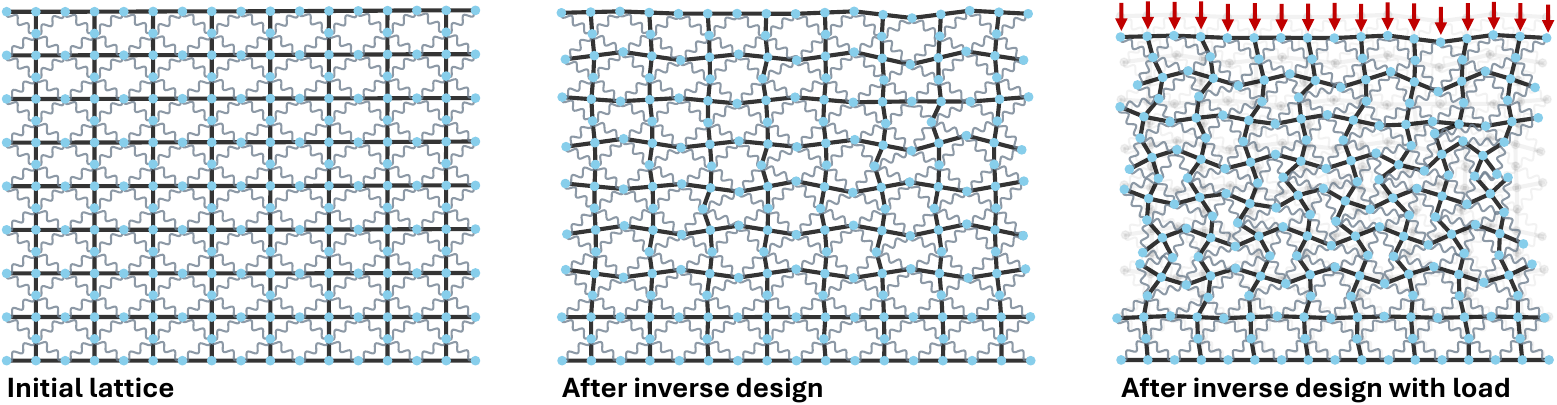}
	\caption{Reconfiguring a $8\times 8$ Totimorphic lattice from a Poisson's ratio of $0$ to $-2$.}\vspace{-4mm}	
	\label{fig:invdesignSI}
\end{figure*}

\begin{figure*}[ht!]
    \centering
    \includegraphics[width=172mm]{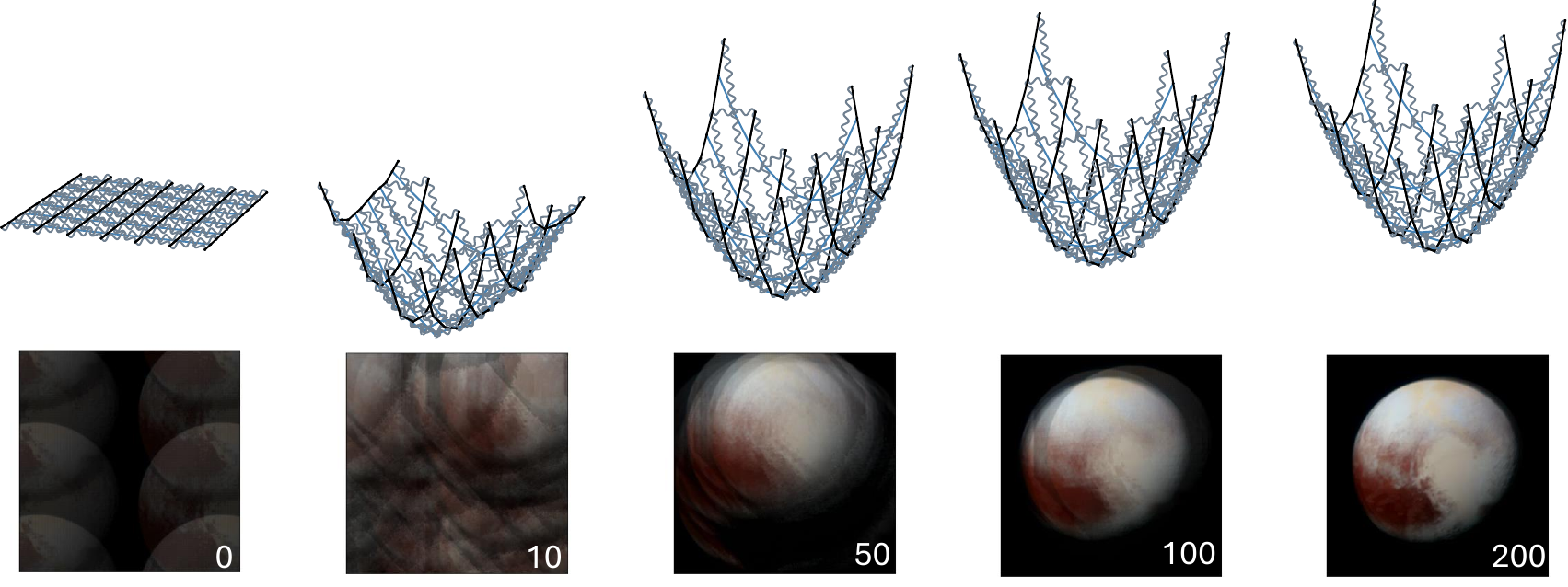}
	\caption{Initial reconfiguration from a flat surface to a curved mirror after 0, 10, 50, 100, and 200 update steps. Both the lattice configuration (top) and an image produced by the mirror at this configuration stage (bottom) are shown.}\vspace{-4mm}	
	\label{fig:telescopeSI}
\end{figure*}

\begin{figure*}[ht!]
    \centering
    \includegraphics[width=0.5\textwidth]{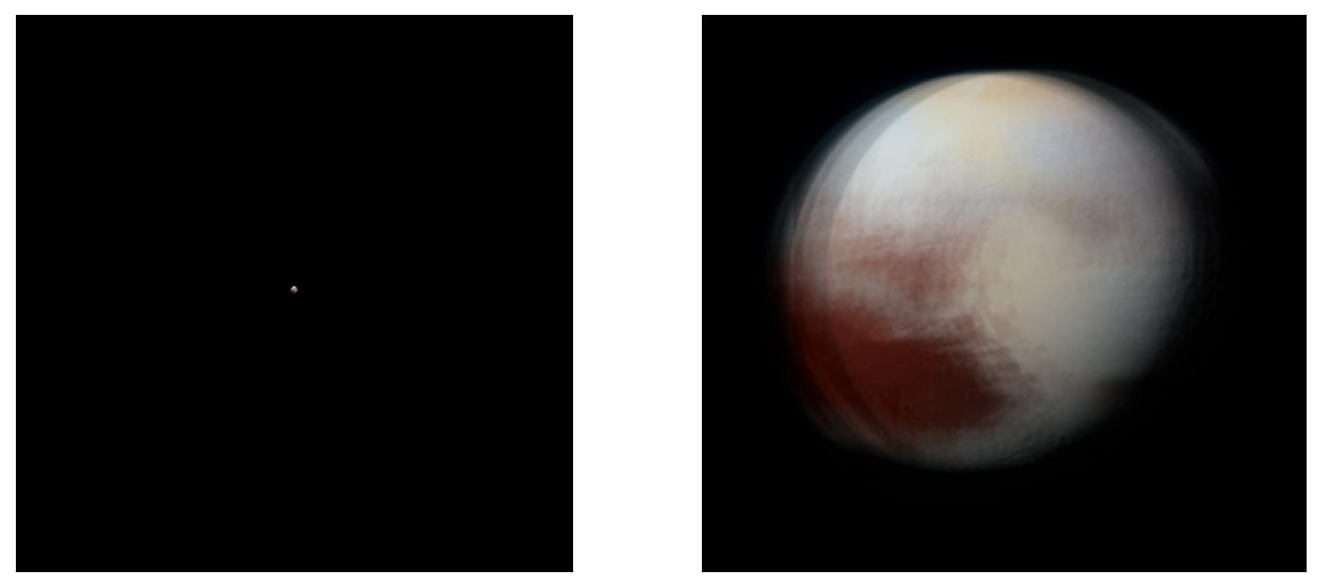}
	\caption{Pluto when imaged under the assumption that it has the same size as the moon and is placed at a distance such that its angular size is equal to the one of Jupiter ($\phi_m \approx 0.01^{\circ}$). The image was obtained with the $20D$ Totimorphic mirror configuration. As can be seen on the left, the image is much smaller due to the  drastically increased distance of the object. Thus, even though the object is still recognizable, sharp details are lost due to the non-perfect focusing of the light (right).}\vspace{-4mm}	
	\label{fig:jupiterSI}
\end{figure*}

\clearpage 

\section*{Supplemental movie captions}

\paragraph{Caption for Movie S1a.}
\textbf{Lattice reconfiguring to attain negative Poisson's ratio.}
Using gradient descent, the Totimorphic lattice reconfigures to attain a Poisson's ratio of $-0.5$.

\paragraph{Caption for Movie S1b.}
\textbf{Lattice reconfiguring to attain positive Poisson's ratio.}
Using gradient descent, the Totimorphic lattice reconfigures to attain a Poisson's ratio of $0.5$.

\paragraph{Caption for Movie S2a.}
\textbf{Reconfigured lattice with negative Poisson's ratio under load.}
Compression experiment with the Totimorphic configuration with Poisson's ratio of $-0.5$.

\paragraph{Caption for Movie S2b.}
\textbf{Reconfigured lattice with positive Poisson's ratio under load.}
Compression experiment with the Totimorphic configuration with Poisson's ratio of $0.5$.

\paragraph{Caption for Movie S3a.}
\textbf{Deployment of a Totimorphic telescope mirror through a series of reconfigurations.}
Initial deployment of the Totimorphic lattice from a collapsed to a flat configuration, which is then reconfigured to focus light at a distance of $20D$, where $D$ is the side length of the lattice.
On the left, a zoom-in of the lattice is shown. The origin of the lattice is also a trainable parameter, which leads to minor movement of the lattice. This could be reduced by increasing the regularisation strength (or choosing the central point in the lattice as the origin and freezing it during optimisation). On the right, the reflected light is shown. The red dot indicates the target focal point.

\paragraph{Caption for Movie S3b.}
\textbf{Adjusting telescope properties through lattice reconfiguration.}
Reconfiguration of the Totimorphic lattice from a focal length of $2D$ to $1D$.
On the left, a zoom-in of the lattice is shown. On the right, the reflected light is shown. The red dot indicates the target focal point.

\end{document}